\documentclass[trackchanges,twocolumn,tighten]{aastex63}

\newcommand{\ltsima}{$\; \buildrel < \over \sim \;$}
\newcommand{\simlt}{\lower.5ex\hbox{\ltsima}}

\def\arcdeg{\hbox{$^\circ$}}
\def\arcmin{\hbox{$^\prime$}}
\def\arcsec{\hbox{$^{\prime\prime}$}}

\usepackage{amsmath}

\shorttitle{NGC~253's Outflow in X-ray}
\shortauthors{Lopez et al.}

\turnoffedit 

\begin{document}

\title{X-Ray Properties of NGC~253's Starburst-Driven Outflow}

\correspondingauthor{Sebastian Lopez}
\email{lopez.764@osu.edu}

\author[0000-0002-2644-0077]{Sebastian Lopez} 
\affil{Department of Astronomy, The Ohio State University, 140 W. 18th Ave., Columbus, OH 43210, USA}
\affil{Center for Cosmology and AstroParticle Physics, The Ohio State University, 191 W. Woodruff Ave., Columbus, OH 43210, USA}

\author[0000-0002-1790-3148]{Laura A. Lopez}
\affil{Department of Astronomy, The Ohio State University, 140 W. 18th Ave., Columbus, OH 43210, USA}
\affil{Center for Cosmology and AstroParticle Physics, The Ohio State University, 191 W. Woodruff Ave., Columbus, OH 43210, USA}
\affil{Flatiron Institute, Center for Computational Astrophysics, NY 10010, USA}

\author[0000-0002-1875-6522]{Dustin D. Nguyen}
\affil{Center for Cosmology and AstroParticle Physics, The Ohio State University, 191 W. Woodruff Ave., Columbus, OH 43210, USA}
\affil{Department of Physics, The Ohio State University, 191 W. Woodruff Ave, Columbus, OH 43210, USA}
\affil{Computational Physics and Methods Group (CCS-2), Los Alamos National Laboratory, Los Alamos, NM 87545, USA}

\author[0000-0003-2377-9574]{Todd A. Thompson}
\affil{Department of Astronomy, The Ohio State University, 140 W. 18th Ave., Columbus, OH 43210, USA}
\affil{Center for Cosmology and AstroParticle Physics, The Ohio State University, 191 W. Woodruff Ave., Columbus, OH 43210, USA}

\author[0000-0002-4822-3559]{Smita Mathur} 
\affil{Department of Astronomy, The Ohio State University, 140 W. 18th Ave., Columbus, OH 43210, USA}
\affil{Center for Cosmology and AstroParticle Physics, The Ohio State University, 191 W. Woodruff Ave., Columbus, OH 43210, USA}

\author[0000-0002-5480-5686]{Alberto D. Bolatto} 
\affil{Department of Astronomy, University of Maryland, College Park, MD 20742, USA}
\affil{Joint Space-Science Institute, University of Maryland, College Park, MD 20742, USA}
\affil{Flatiron Institute, Center for Computational Astrophysics, NY 10010, USA}
\affil{Visiting Astronomer, National Radio Astronomy Observatory, VA 22903, USA}

\author[0000-0001-7855-8336]{Neven Vulic}
\affil{Eureka Scientific, Inc., 2452 Delmer Street, Suite 100, Oakland, CA 94602-3017, USA}
\affil{NASA Goddard Space Flight Center, Code 662, Greenbelt, MD 20771, USA}
\affil{Department of Astronomy, University of Maryland, College Park, MD 20742-2421, USA}
\affil{Center for Research and Exploration in Space Science and Technology, NASA/GSFC, Greenbelt, MD 20771, USA}

\author[0000-0002-5783-145X]{Amy Sardone}
\affil{Department of Astronomy, The Ohio State University, 140 W. 18th Ave., Columbus, OH 43210, USA}
\affil{Center for Cosmology and AstroParticle Physics, The Ohio State University, 191 W. Woodruff Ave., Columbus, OH 43210, USA}

\begin{abstract}

We analyze image and spectral data from $\approx$365~ks of observations from the {\it Chandra} X-ray Observatory of the nearby, edge-on starburst galaxy NGC~253 to constrain properties of the hot phase of the outflow. We focus our analysis on the $-$1.1 to $+$0.63 kpc region of the outflow and define several regions for spectral extraction where we determine best-fit temperatures and metal abundances. We find that the temperatures and electron densities peak in the central $\sim$250 pc region of the outflow and decrease with distance. These temperature and density profiles are in disagreement with an adiabatic spherically expanding starburst wind model and suggest the presence of additional physics such as mass loading and non-spherical outflow geometry. Our derived temperatures and densities yield few-Myr cooling times in the nuclear region, which  may imply that the hot gas can undergo bulk radiative cooling as it escapes along the minor axis. Our metal abundances of O, Ne, Mg, Si, S, and Fe all peak in the central region and decrease with distance along the outflow, with the exception of Ne which maintains a flat distribution. The metal abundances indicate significant dilution outside of the starburst region. We also find estimates on the mass outflow rates which are $2.8\:M_{\odot}/\rm{yr}$ in the northern outflow and $3.2\:M_{\odot}/\rm{yr}$ in the southern outflow. Additionally, we detect emission from charge exchange and find it has a significant contribution ($20-42$\%) to the total broad-band ($0.5-7$~keV) X-ray emission in the central and southern regions of the outflow. 

\end{abstract}


\section{Introduction} \label{sec:intro}

Starburst galaxies are characterized by their prominent multiphase outflows known as galactic winds \citep{Veilleux2005,Rubin2014,Veilleux2020} and are the result of an intense period of star formation. These galactic winds have important effects on their host galaxies where they alter the metal content of the disk \citep{Peeples2011,Finlator2008} and are able to enrich the surrounding circumgalactic medium (CGM) and intergalactic medium (IGM) \citep{Fielding2017,Opp2008}. 

NGC 253 is a nearby (3.5 Mpc; \citealt{Rekola2005}), edge-on ($i=76\arcdeg{}$; \citealt{McCormick2013}) starburst which has a multiphase outflow that has been well-studied across the electromagnetic spectrum: e.g., the $\approx10^{2}$~K gas at mm wavelengths \citep{Bolatto2013,Leroy2015,Krieger2019}, the $\approx10^4$ K gas at optical wavelengths \citep{Westmoquette2011}, and the $\approx 10^7$ K gas at X-ray wavelengths \citep{Strickland2000,Strickland2002,Bauer2007,Mitsuishi2013,Wik2014}. 

Previous work on the hot gas component has found temperature, column densities, and metallicity constraints of NGC~253's outflow. \cite{Strickland2000} measured temperatures and column densities in five different regions along the outflow by fitting a single temperature plasma model using data from \textit{Chandra}. \cite{Bauer2007} measured temperatures along the outflow using emission line ratios from {\it XMM-Newton} spectra. \cite{Mitsuishi2013} measured both metallicities and temperatures of the disk, superwind, and halo region using {\it XMM-Newton} and {\it Suzaku} data. Missing from previous work is an X-ray study of the outflow that incorporates both the coverage and resolution along the outflow of \cite{Strickland2000} and \cite{Bauer2007}, with the more sophisticated model and metallicity constraints of \cite{Mitsuishi2013}. 

Also missing from previous X-ray analyses of NGC 253's outflow is the inclusion of charge exchange (CX) in spectral models used to constrain the temperatures and metal abundances. CX is the stripping of an electron from a neutral atom by an ion. CX has been found to produce X-ray emission in galactic winds \citep{Wang2012,Liu2012} and recently, \cite{Lopez2020} found that charge exchange contributes between 8\%–25\% of the total absorption-corrected, broadband ($0.5–7$ keV) X-ray flux in the starburst M82. As a result, the absence of CX in spectral models can lead to inaccuracies in constraints on the outflow properties. 

In this paper we follow the methodology employed in \cite{Lopez2020} to constrain temperature, metal abundances, column density, and number density of NGC 253's outflow using \textit{Chandra} images and spectra. We employ recent spectral models that include CX to account for its contribution to the line emission. In Section~\ref{sec:methods}, we describe the \textit{Chandra} observations used and provide details for the procedures used to produce X-ray images and spectra of NGC 253. In Section~\ref{sec:results}, we present the results of our spectral fitting along the outflow as well as across the disk. In Section~\ref{sec:disc}, we compare our work to previous X-ray studies of NGC 253, to the results of \cite{Lopez2020} on M82, and to the predictions of galactic wind models. In this section, we also derive mass outflow rate estimates and quantify the contribution of CX to the broadband emission. In Section~\ref{sec:con}, we summarize our findings and outline future work to be done to constrain the hot phase in outflows. 

Throughout the paper we assume a distance of 3.5 Mpc to NGC 253 \citep{Rekola2005}, where 1\arcmin{}=1.0 kpc.

\begin{figure}
    \centering
    \includegraphics[width=0.45\textwidth]{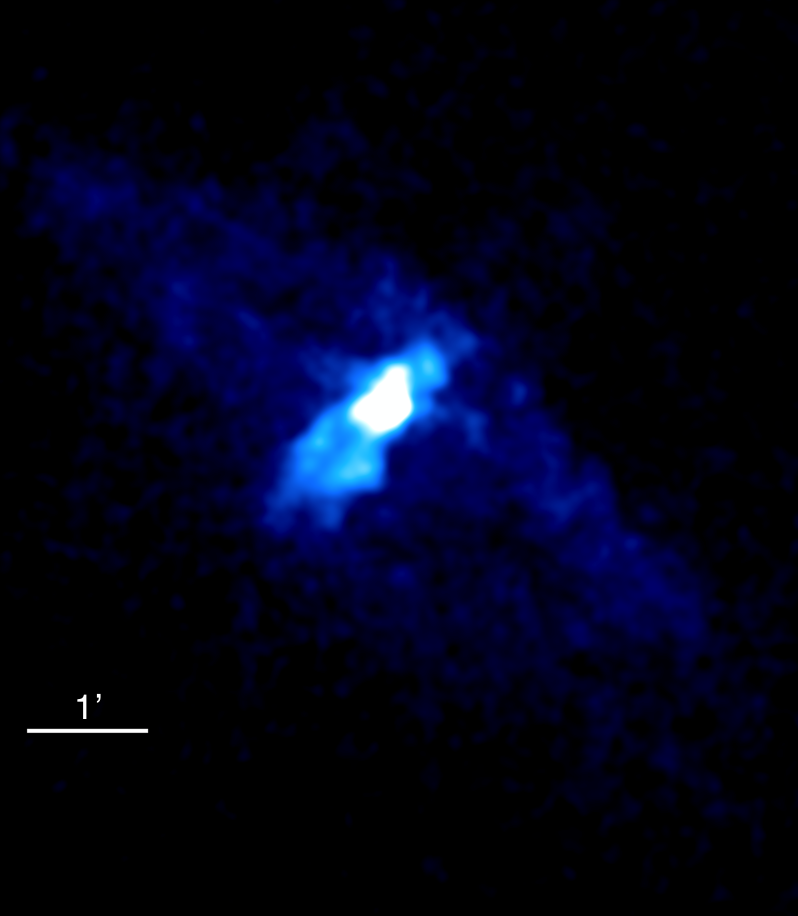}
    \caption{Broad-band ($0.5-7.0$~keV) X-ray image of NGC 253. The 1\arcmin\ label corresponds to a physical size of 1.0 kpc. In the image North is up and East is left. The $D_{25}$ ellipse of NGC~253 is 27.5\arcmin \citep{RC3} and extends beyond this image.}
    \label{fig:ngc253_xray}
\end{figure}

\begin{deluxetable}{lrc}
\tablecolumns{3}
\tablewidth{0pt} \tablecaption{{\it Chandra} Observations \label{table:data}} 
\tablehead{\colhead{ObsID} & \colhead{Exposure} & \colhead{UT Start Date}}  
\startdata
790 & 45~ks & 1999-12-27 \\
969 & 15~ks & 1999-12-16  \\
3931 & 85~ks & 2003-09-19  \\
13830 & 20~ks & 2012-09-02  \\
13831 & 20~ks & 2012-09-18 \\
13832 & 20~ks & 2012-11-16 \\
20343 & 160~ks & 2018-08-15  
\enddata
\end{deluxetable}

\begin{figure}
    \centering
    \includegraphics[width=0.45\textwidth]{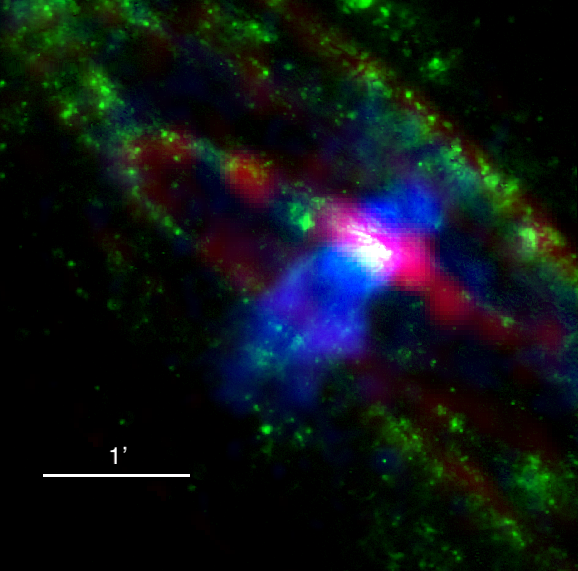}
    \caption{Three color image of NGC 253, where blue is broad-band (0.5-7 keV) {\it Chandra} X-ray, green is H$\alpha$ \citep{LH1995}, and red is CO~($2-1$) \citep{Leroy2021} emission. The 1\arcmin{} label corresponds to 1.0 kpc, and in the image, North is up and East is left.}
    \label{fig:color}
\end{figure}

\section{Methods} \label{sec:methods}

NGC~253 was observed seven times with {\it Chandra} from 1999 to 2018, as detailed in Table~\ref{table:data}. The first three observations used the ACIS-S array, while the rest used ACIS-I.  

Data were downloaded from the archive and reduced using the Chandra Interactive Analysis of Observations {\sc ciao} version 4.14 \citep{CIAO2006}. Using the \textit{merge\_obs} function, the observations were combined, and the \textit{wavdetect} function identified point sources in the image that were removed with the \textit{dmfilth} command. \edit1{The detection sensitivity for the removed point sources was $7.14\times10^{-16}$ ergs/cm$^2$/s. The sensitivity was calculated using WebPIMMS\footnote{https://heasarc.gsfc.nasa.gov/cgi-bin/Tools/w3pimms/w3pimms.pl}, assuming a point source with a power-law spectrum of photon index 1.7.} Each of the observations were also checked for background flares and none were found.

The final, broad-band ($0.5-7.0$~keV) X-ray image of the diffuse emission is shown in Figure~\ref{fig:ngc253_xray}. From this image, the outflow extends $\approx$0.63\arcmin\ to the north and $\approx$1.1\arcmin\ south of the starburst. Additionally, the galactic disk is evident in the diffuse X-rays as well, following the spiral structure as seen in the three-color image with CO \citep{Leroy2021} and H$\alpha$ \citep{LH1995} shown in Figure~\ref{fig:color}.

\subsection{Outflow Spectral Analysis}

\begin{figure*}
    \centering
    \includegraphics[width=\textwidth]{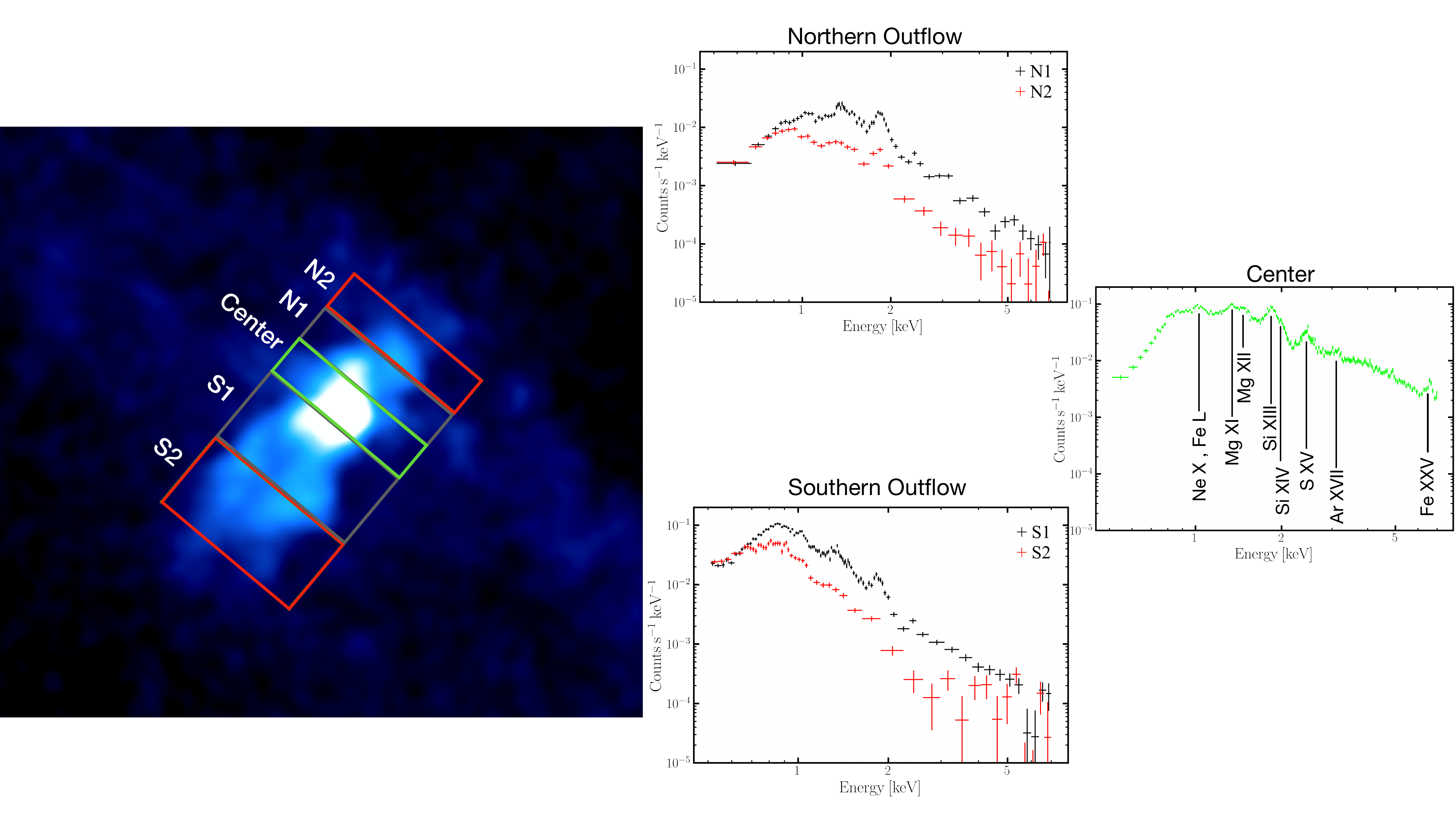} 
    \caption{\textit{Left:} A zoom-in of Figure~\ref{fig:ngc253_xray} with 5 regions marked where spectra were extracted. The center and northern (N1 and N2) regions are 0.25\arcmin{} in height and 1\arcmin{} in width. The southern regions (S1 and S2) are 0.5\arcmin{} in height and 1\arcmin{} in width. \textit{Right:} Background-subtracted spectra from the center region and from each hemisphere with the emission lines labeled. The colors of each region correspond to the same colors on the spectra plots. Metals are found in every region. The Fe~{\sc xxv}, Ar~{\sc xvii}, and S~{\sc xv} lines were only detected in the center region, indicating a hotter plasma there.}
    \label{fig:ngc253_box}
\end{figure*}

In order to constrain the properties from the hot gas outflow, spectra were extracted using {\sc ciao}. We defined several regions along the minor axis of the outflow as a function of distance from the kinematic center of NGC~253 \citep{Muller-Sanchez2010}.
In Figure~\ref{fig:ngc253_box}, the areas where spectra were extracted are shown. From the center to the north of the outflow, the regions were 0.25\arcmin\ in height and 1\arcmin\ in width, while the southern regions were 0.5\arcmin\ in height and 1\arcmin\ in width. The region sizes were set to achieve $\gtrsim5000$ counts per region (in order to reliably constrain the metal abundances) and were placed qualitatively in different sections of the outflow (the starburst nucleus, north, and south). 

Background regions were defined for each of the observations and subtracted from each of the source regions. The background regions totaled an area of 32~arcmin$^{2}$. These regions were placed far from the disk in order to avoid diffuse emission from NGC~253, while also being located on the same ACIS chip as the source regions. The background regions are $\geq$2.5\arcmin\ from the center of the galaxy. 

Once extracted and background subtracted, the spectra were modeled with XSPEC \citep{XSPEC} Version 12.12. The XSPEC model that was used to fit the spectral data differed depending on the region. All the models included a multiplicative constant component (\textsc{const}), two absorption components (\textsc{phabs*phabs}), a powerlaw component (\textsc{powerlaw}), and at least one optically thin, thermal plasma component (\textsc{vapec}). The \textsc{const} component was allowed to vary and accounted for variations in emission between the observations. The first \textsc{phabs} component accounted for the galactic absorption in the direction of NGC 253, $N_{\rm H}^{\rm MW}=1.4\times 10^{20}\:\mathrm{cm^{-2}}$ \citep{DL1990} and was frozen. The second component $N_{\rm H}^{\rm NGC~253}$ represented the intrinsic absorption of NGC~253 and was allowed to vary. \edit1{Both \textsc{phabs} components have their metal abundance set to solar, consistent with observations of the NGC 253 nuclear region \citep{Mills2021,Beck2022}.} The \textsc{powerlaw} component accounted for the non-thermal X-ray emission from e.g., X-ray binaries, and the \textsc{vapec} component represented the thermal plasma with variable abundances and temperatures. We adopted solar abundances from \cite{Asplund2009} and photoionization cross sections from \cite{Verner1996}.

Several other components were added as necessary to improve fits. The center region required a second \textsc{vapec} component that accounted for a hotter thermal plasma component. The center, S1, and S2 regions also included an AtomDB\footnote{http://www.atomdb.org/CX/} charge-exchange (CX) component \textsc{vacx} that accounted for the emission from electrons captured by ions from neutral atoms~\citep{AtomDB}. \edit1{The \textsc{vacx} and additional \textsc{vapec} components have their metal abundances tied to the first \textsc{vapec} component.} Regions N1 and N2 were found to not need the \textsc{vacx} component based on F-tests. 

The final model used for regions N1 and N2 was \textsc{const*phabs*phabs*(vapec+powerlaw)}. For regions S1 and S2, the model was \textsc{const*phabs*phabs*(vapec+powerlaw+vacx)}, and for the center region, the model was \textsc{const*phabs*phabs*(vapec+vapec+powerlaw+vacx)}.

\subsection{Disk Spectral Analysis}
X-ray emission was also evident in the disk of NGC~253. To constrain its properties, we extracted and modeled spectra from nine 1\arcmin{} by 1\arcmin{} regions shown in Figure~\ref{fig:disk_reg}. The disk spectral models were similar to those of the outflow except the abundances were not allowed to vary individually because there were not enough counts to reliably constrain the abundances. \edit1{Instead, the total metallicity of the \textsc{apec} component, $Z$, was set to solar to be consistent with observations \citep{Mills2021,Beck2022}}. The XSPEC model for the disk was \textsc{const*phabs*phabs*(apec+powerlaw)}.

\section{Results} \label{sec:results}

\subsection{Outflow Spectral Results} \label{subsec:spectralresults}

In Figure~\ref{fig:ngc253_box}, we present the X-ray spectra extracted from the outflow. We find several emission lines in these regions as identified in the center region spectrum. We detect emission lines from Ne, Fe, Mg, Si, S, and Ar. The Fe~{\sc xxv}, Ar~{\sc xvii}, and S~{\sc xv} lines are only apparent in the central region because of the hotter plasma located there. 

\begin{deluxetable*}{lcccccccccr}
\tablecolumns{11}
\tablewidth{0pt} \tablecaption{Spectral Fit Results\tablenotemark{a} \label{table:fitresults}} 
\tablehead{\colhead{Reg.} & \colhead{$r$} & \colhead{$N_{\rm H}^{\rm NGC253}$} & \colhead{$kT$} &  \colhead{O/O$_{\sun}$} & \colhead{Ne/Ne$_{\sun}$} & \colhead{Mg/Mg$_{\sun}$} & \colhead{Si/Si$_{\sun}$} & \colhead{S/S$_{\sun}$} & \colhead{Fe/Fe$_{\sun}$} & \colhead{$\chi^{2}$/d.o.f.} \\
\colhead{} & \colhead{(kpc)} & \colhead{($\times10^{22}$~cm$^{-2}$)} & \colhead{(keV)} & \colhead{} & \colhead{} & \colhead{} & \colhead{} & \colhead{} & \colhead{}
}  
\startdata
N2 & 0.52 & 0.17$\pm$0.09 & 0.71$\pm$0.08 &  $<$0.96 & $<$0.51 & $0.45_{-0.18}^{+0.45}$ & $1.0_{-0.27}^{+0.58}$ & $1.0_{-0.28}^{+0.76}$ & $0.09_{-0.03}^{+0.06}$ & $240/227$ \\
N1 &  0.26 & $0.66_{-0.09}^{+0.12}$ & $0.73\pm0.04$ & 1 & 1 & $1.0_{-0.17}^{+0.20}$ & $1.4_{-0.20}^{+0.23}$ & $1.4_{-0.39}^{+0.44}$ & $0.05_{-0.03}^{+0.04}$ & $551/498$ \\
Center\tablenotemark{b} & 0 & 0.86$\pm$0.09 & 0.98$\pm$0.02 & $1.3_{-0.57}^{+0.97}$ & $0.89_{-0.30}^{+0.50}$ & $1.8_{-0.55}^{+0.92}$ & $4.0_{-1.1}^{+1.9}$ & $6.6_{-1.8}^{+3.2}$ & $2.7_{-0.74}^{+1.4}$ & $1979/1552$ \\
S1 & $-$0.39 & 0.15$\pm$0.04 & 0.65$\pm$0.03 & $0.20_{-0.08}^{+0.12}$ & $0.69_{-0.20}^{+0.27}$ & $0.49_{-0.12}^{+0.15}$ & $0.56_{-0.11}^{+0.14}$ & $0.66_{-0.28}^{+0.32}$ & 0.26$\pm$0.04 & $752/636$\\
S2 & $-$0.92 & $0.18_{-0.09}^{+0.10}$ & 0.37$\pm$0.04 & $0.24_{-0.10}^{+0.15}$ & $0.60_{-0.21}^{+0.31}$ & $0.35_{-0.13}^{+0.19}$ & 1 & 1 & $0.21_{-0.06}^{+0.08}$ & $272/258$ \\
\enddata
\tablenotetext{a}{Abundances with values of 1 are frozen to solar values in the fits.}
\tablenotetext{b}{The temperature for the second hotter thermal plasma components is $kT_{2} = 5.5_{-0.8}^{+1.2}$ keV.}

\vspace{5mm}
\end{deluxetable*}

\begin{figure*}
    \centering
    \includegraphics[width=\textwidth]{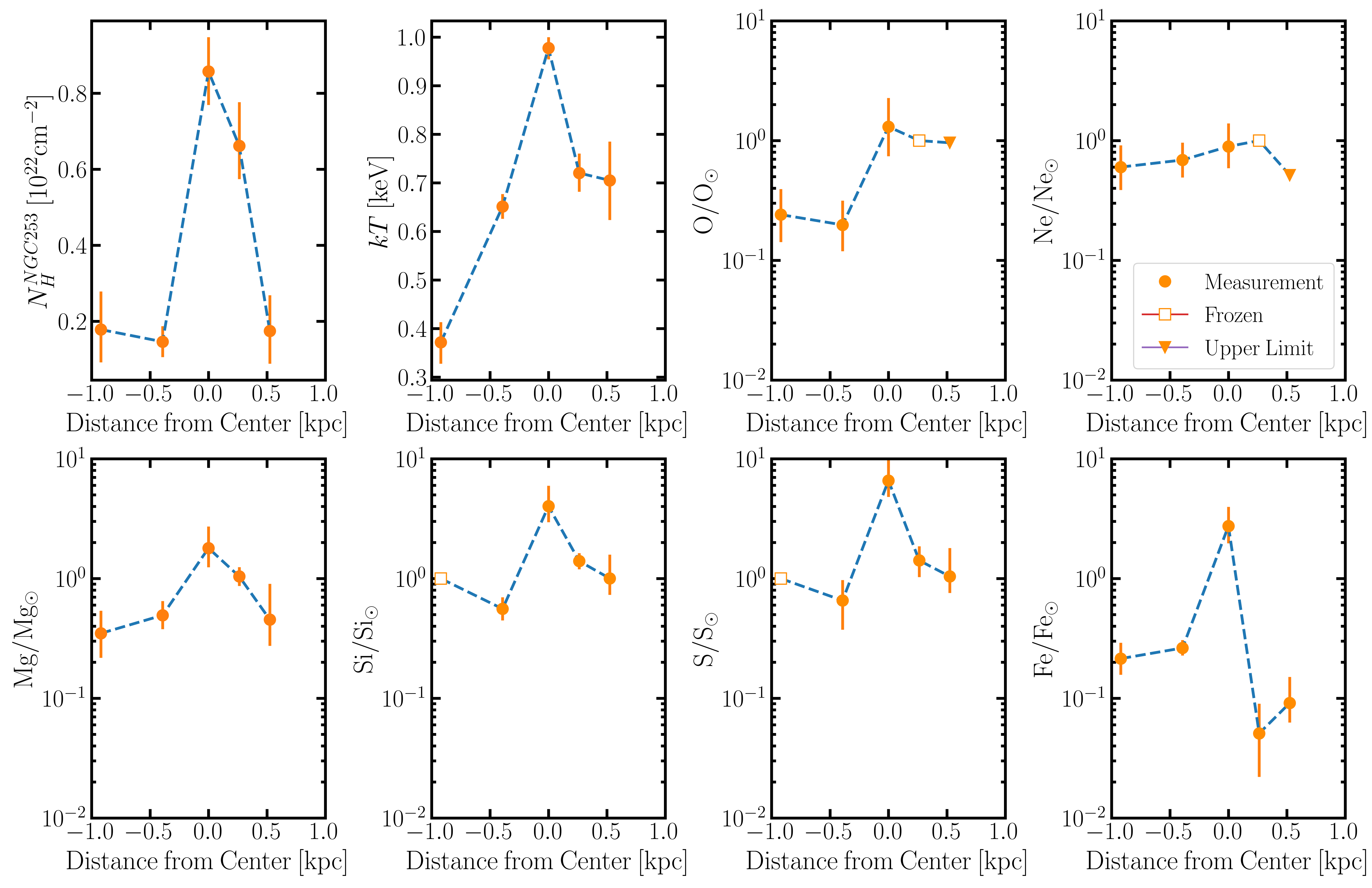}
    \caption{Best-fit parameters plotted as a function of the distance along minor axis of NGC~253. The filled circles are measurements, open circles are frozen values, and upside down triangles are upper limits. $kT$ and $N^{\rm NGC253}_{\rm H}$ peak in the center and decrease with distance. With the exception of Ne, the metal abundances also peak in the center and decrease with distance. The metal distributions indicate much dilution of the gas outside of the starburst region.}
    \label{fig:abund}
\end{figure*}

Using the models described in Section~\ref{sec:methods} and the extracted spectra, we find the best-fit values of column density $N^{\rm NGC253}_{\rm H}$, temperature $kT$, and metal abundances. We report these values in Table~\ref{table:fitresults} and plot them in Figure~\ref{fig:abund}. 
\edit1{We find that the models fit the data well, with the reduced $\chi^2$ values in each region being near one, and the largest being the center region with a reduced $\chi^2$ of 1.28 and a null hypothesis of $7.67\times10^{-13}$.} 

All the best-fit values, with the exception of Ne, have distributions that peak in the center and decrease outward along the outflow to the north and south. For $N^{\rm NGC253}_{\rm H}$ the gradient is similar in the northern and southern outflow, with a peak of $N^{\rm NGC253}_{\rm H} = (8.6\pm0.9)\times10^{21}$~cm$^{-2}$ in the center and decreasing to $N^{\rm NGC253}_{\rm H}$ $\approx (1-2)\times10^{21}$~cm$^{-2}$. For $kT$, it peaks at the center, with $kT = 0.98\pm0.02$~keV and decreases going outward along the outflow axis. We note that the second, hotter component in the center had a best-fit value of $kT_{2} = 5.5^{+1.2}_{-0.8}$~keV.

The metal abundances also peak in the center, with S having the highest value of 6.6$^{+3.2}_{-1.8}$ $\rm{S/S_{\odot}}$ and Ne having the smallest peak with a relatively flat distribution within the uncertainties. \edit1{We note in Section~\ref{subsec:M82comp} that the high S value may be due to contributions from the second temperature component.} The profiles show that there is little enhancement of the metals beyond the central region, with the outer regions containing metal abundances of around the solar value or lower. The Fe abundance is particularly low, with all the outflow regions below $\lesssim$0.30$ \rm{Fe/Fe_{\odot}}$. The decrease in abundances along the outflow may suggest mass loading or indicate other physics should be considered, such as non-equilibrium ionization.

\begin{deluxetable*}{lccccccccc} \rotate
\tablecolumns{7}
\tablewidth{0pt} \tablecaption{Physical Parameters of the Center and Outflow Regions \label{table:ne}} 
\tablehead{\colhead{Reg.} & \colhead{$norm$} & \colhead{$L_X$} & $L_{CX}/L_{X}$& \colhead{$EM$} & \colhead{$R$\tablenotemark{a}} & \colhead{$V$\tablenotemark{a}} & \colhead{$n_{\rm e}$\tablenotemark{a}} & \colhead{$P/k$\tablenotemark{a}} & \colhead{$t_{\rm cool}$\tablenotemark{a}}  \\
\colhead{} & \colhead{} & \colhead{($\times 10^{38}$erg/s)} & \colhead{} & \colhead{($\times10^{62}$ cm$^{-3}$)} & \colhead{($\times10^{21}$ cm)} & \colhead{($\times10^{63}$~cm$^{3}$)} & \colhead{(cm$^{-3}$)}  & \colhead{($\times10^{6}$ K~cm$^{-3}$)} & \colhead{(Myr)}}  
\startdata
N2 & 1.5$\times10^{-4}$ & 1.1& - & 0.32 & 1.5 (0.68)& 5.3 (3.0)& $0.07_{-0.02}^{+0.04}$ ($0.09_{-0.03}^{+0.05}$)& $1.2_{-0.33}^{+0.67}$ ($1.5_{-0.43}^{+0.88}$)& $48_{-13}^{+28}$ ($36_{-10}^{+21}$)  \\
N1 & 6.3$\times10^{-4}$ & 5.2& - & 1.6& 1.3 (0.58)& 3.9 (2.2)& $0.17\pm0.02$ ($0.22\pm0.03$)& $2.8_{-0.33}^{+0.41}$ ($3.8_{-0.44}^{+0.55}$)& $21_{-2.4}^{+3.0}$ ($15_{-1.8}^{+2.3}$) \\
Center\tablenotemark{b} & 6.6$\times10^{-4}$ & 47& 0.42 & 1.4& 0.73 (0.26) & 1.3 (0.59)& $0.30\pm0.02$ ($0.44_{-0.03}^{+0.04}$)& $6.7_{-0.52}^{+0.56}$ ($10_{-0.78}^{+0.84}$)& $17_{-1.4}^{+1.5}$ ($12_{-0.91}^{+0.97}$)\\
S1 & 7.1$\times10^{-4}$ & 7.7& 0.20 & 1.5& 1.3 (0.63)& 7.8 (4.7)& $0.13_{-0.01}^{+0.02}$ ($0.16\pm0.02$)& $1.9_{-0.20}^{+0.24}$ ($2.5_{-0.26}^{+0.30}$)& $25_{-2.6}^{+3.1}$ ($19_{-2.0}^{+2.4}$)\\ 
S2 & 2.5$\times10^{-4}$ & 3.3& 0.33 & 0.53 & 1.4 (0.73)& 9.1 (6.0)& $0.07_{-0.01}^{+0.02}$ ($0.09\pm0.02$)& $0.60_{-0.12}^{+0.16}$ ($0.74_{-0.15}^{+0.20}$)& $27_{-5.2}^{+7.1}$ ($21_{-4.2}^{+5.7}$) \\
\enddata
\tablenotetext{a}{The values listed for $R$, $V$, $n_{\rm e}$, $P/k$, and $t_{cool}$ are for the 95\% emission radius. The values in parentheses are for the 68\% emission radius (see Section~\ref{subsec:spectralresults} for details of these measurements).}
\tablenotetext{b}{The parameters of the hotter component $kT_{2}$ in center region are: $norm_{2} = 2.7\times10^{-4}$, $n_{\rm e,2} = 0.19\pm0.02$~cm$^{-3}$, $P_{2}/k = 2.4\pm0.2\times10^{7}$~K~cm$^{-3}$, and $t_{\rm cool,2} = 211_{-19}^{+21}$~Myr. For the 68\% emission radius, the values are: $n_{\rm e,2} = 0.28\pm0.03$~cm$^{-3}$, $P_{2}/k = 3.6_{-0.34}^{+0.36}\times10^{7}$~K~cm$^{-3}$, and $t_{\rm cool,2} = 140_{-13}^{+14}$~Myr.}

\vspace{-5mm}
\end{deluxetable*}

From the best-fit values, we calculate several other properties of the outflow for each region and present them in Table~\ref{table:ne}. The norm is defined as $norm = (10^{-14}EM)/4\pi D^2$, where the emission measure is $EM = \int n_{\rm e} n_{\rm H} dV$. By setting $n_{\rm e}=1.2n_{\rm H}$, we calculate $n_{\rm e}$ as $n_{\rm e} = (1.5\times10^{15}normD^2/fV)^{1/2}$ where $f$ is the filling factor and $V$ is the volume. In computing $V$, we assume that each region has a cylindrical volume of radius $R$ that we measure using broad-band, X-ray brightness profiles. We make 2\arcsec\ slits along the major axis to create the brightness profile, and we define $R$ as the scale that encompasses 95\% of the X-ray brightness (for reference, we also list the values corresponding to the 68\% X-ray brightness in Table~\ref{table:ne} as well). \edit1{We note that the measurement of $R$ is influenced by our choice of regions over which we extracted the profiles. We found that if we define regions larger than the ones used in Section 2.1, the 95\% emission radii were overestimated due to more background emission being included.}

We assume a filling factor of $f = 1$, though we note that due to processes like mass loading, the filling factor may be lower. The other measured quantities are the thermal pressure $P/k=2n_{\rm e}T$ and the cooling time $t_{\rm cool} = 3kT/\Lambda n_{\rm e}$.  We find the radiative cooling function $\Lambda$ using CHIANTI ~\citep{CHIANTI} at solar metallicity assuming a thin, thermal plasma. 

The results in Table~\ref{table:ne} show that $R$, and consequently the volume $V$, increases with distance from the center of the galaxy. $n_{\rm e}$ peaks in the center region with a value of $n_{\rm e} = 0.30\:\rm{cm^{-3}}$ and decreases along the outflow. The pressure also peaks in the center region with a value of $6.7\times10^6\rm{\:K\:cm^{-3}}$. The cooling time has the lowest value of $t_{\rm cool} = 17$ Myr in the central region, and the largest value of \edit1{48} Myr in region N2. S1, and S2 have comparable cooling times of $25-27$~Myr. 

As noted in Section~\ref{sec:model_compare}, when using smaller regions, the cooling times are much shorter (the shortest time is $t_{\rm cool} = 1.26$ Myr). The larger region cooling times appear to overestimate those of the smaller regions. 

We also calculate the same physical quantities shown in Table~\ref{table:ne} for the hotter temperature component, $kT_2$, that is only present in the center region. 
We find that the hotter component has a lower $n_{\rm e}$ of $0.19\:\rm{cm^{-3}}$ than the cooler component.
The hotter component has a higher pressure of $2.4\times10^7\rm{\:K\:cm^{-3}}$ and a longer cooling time of 211 Myr than the respective values in the cooler component.

\subsection{Disk Spectral Results}

\begin{figure*}
    \centering
    \includegraphics[width=0.40\textwidth]{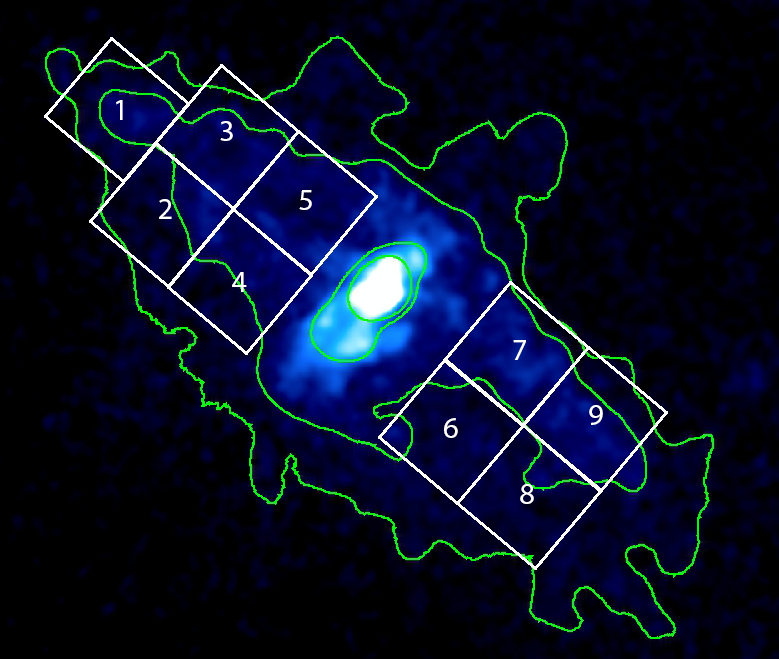}
    \includegraphics[width=0.55\textwidth]{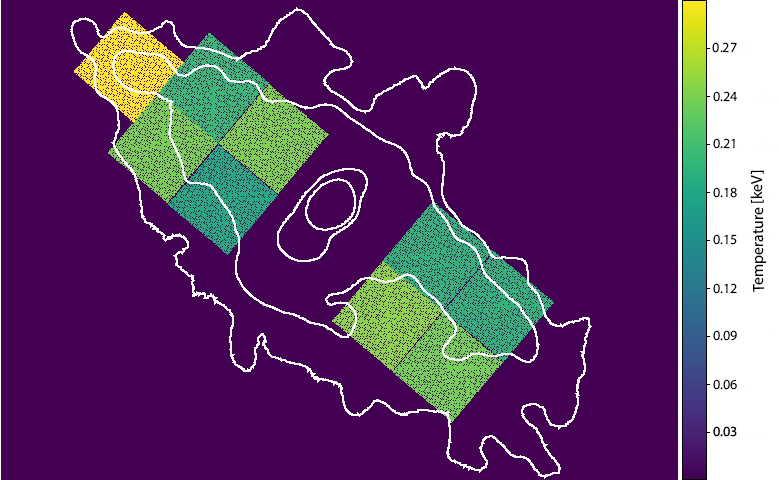}
    \caption{\textit{Left:} Broad-band X-ray image of NGC 253 with contours of the 0.5-7 keV emission overplotted along with numbered regions used for spectral extraction from the disk. Each region is 1\arcmin{} by 1\arcmin. \textit{Right:} NGC~253 broad-band X-ray contours along with the regions of spectral extraction each filled with a best-fit $kT$ value in keV.}
    \label{fig:disk_reg}
\end{figure*}

Using the spectral model described in Section \ref{sec:methods} we measure $kT$ and $N^{\rm NGC253}_{\rm H}$ for the diffuse X-ray emission in the disk (Table~\ref{table:disk_val}). We find that the $kT$ values are in the range of 0.18 $-$ 0.30 keV, and the $N^{\rm NGC253}_{\rm H}$ values are in the range of (0.31 - 0.72)$\mathrm{\times 10^{22} cm^{-2}}$. As shown in the right panel of  Figure~\ref{fig:disk_reg}, we find the highest $kT$ in Region 1 and the lowest in Region 4. 
 
\begin{deluxetable}{lccclccc}
\tablecolumns{4}
\tablewidth{0pt} \tablecaption{Disk Spectral Results \label{table:disk_val}} 
\tablehead{\colhead{Reg.} & \colhead{$N_{\rm H}^{\rm NGC253}$} & \colhead{$kT$}   & \colhead{$\chi^{2}$/d.o.f.} \\
\colhead{} & \colhead{($\times10^{22}\:\rm{cm^{-2}}$)} & \colhead{(keV)} &  \colhead{}}  
\startdata
1 &	$0.40_{-0.14}^{+0.19}$ &	$0.30_{-0.06}^{+0.04}$ &	$212/185$ \\
2 &	$0.37_{-0.15}^{+0.16}$ &	$0.24_{-0.02}^{+0.05}$ &	$186/178$ \\
3 &	$0.72_{-0.19}^{+0.08}$ &	$0.20_{-0.02}^{+0.05}$ &	$206/199$ \\
4 &	$0.71_{-0.11}^{+0.10}$ &	$0.18_{-0.01}^{+0.02}$ &	$191/181$ \\
5 &	$0.54_{-0.10}^{+0.09}$ &	$0.24_{-0.01}^{+0.02}$ &	$281/237$ \\
6 &	$<0.53$                &	$0.25_{-0.03}^{+0.11}$ &	$279/273$ \\
7 &	$0.69_{-0.06}^{+0.05}$ &	$0.20_{-0.01}^{+0.01}$ &	$396/335$ \\
8 &	$0.31_{-0.20}^{+0.29}$ &	$0.24_{-0.05}^{+0.03}$ &	$225/243$ \\
9 &	$0.64_{-0.07}^{+0.07}$ &	$0.20_{-0.01}^{+0.02}$ &	$349/303$ \\
\enddata
\vspace{-5mm}
\end{deluxetable}

\section{Discussion} \label{sec:disc}

\subsection{Comparison to previous work on NGC 253}
Previous X-ray studies of NGC~253 have been performed and are complementary to our analysis. In \cite{Strickland2000}, 15 ks of {\it Chandra} X-ray observations were used to study NGC~253's outflow, much shallower than the 365 ks data considered in this work.

In their work, X-ray spectra were extracted along the starburst outflow from regions spanning a similar area as the ones in this paper. There were five regions each of which measured about 0.33\arcmin{} in height and 0.8\arcmin{} in width. The spectra were fitted using a single-temperature plasma model, where the $kT$, $N^{\rm NGC253}_{\rm H}$, and normalization were allowed to vary. In their analysis, the model was only fit to the soft X-rays ($0.3-3$~keV) due to near zero count rates at harder X-ray energies in most regions.

\cite{Strickland2000} found that the $N^{\rm NGC253}_{\rm H}$ values in the southern hemisphere range from $(0.034-0.056)\times10^{22}\:\mathrm{cm^{-2}}$ and $kT$ values range from $0.46-0.63$ keV. In the northern outflow, their $N^{\rm NGC253}_{\rm H}$ values range from $(0.473-0.922)\times10^{22}\:\mathrm{cm^{-2}}$ and their temperatures range from $0.59-0.66$ keV. 

Our $kT$ values in the southern outflow are in a similar range ($0.37-0.65$ keV); however our $N^{\rm NGC253}_{\rm H}$ values differ greatly ($(0.15-0.18)\times10^{22}\:\mathrm{cm^{-2}}$). We find the opposite when comparing our northern outflow values, where the  $N^{\rm NGC253}_{\rm H}$ values are comparable with ours ($(0.17-0.66)\times10^{22}\:\mathrm{cm^{-2}}$), but the $kT$ results are discrepant ($0.71-0.72$ keV). 

\edit1{The difference between our derived values and those of \cite{Strickland2000} may result from their limited energy range (0.3-3 keV) and differences in the spectral model they adopted.} As a result of \cite{Strickland2000} using a softer X-ray range, their derived temperatures are lower in regions of higher $N^{\rm NGC253}_{\rm H}$ (the northern outflow), where softer X-rays are attenuated. By contrast, in the southern outflow where $N_{\rm H}^{\rm NGC253}$ is lower, we find comparable $kT$ values. If we adopt the energy range of \cite{Strickland2000}, then we find lower temperatures in the northern outflow ($0.66\pm{0.04}$ and $0.71^{+0.08}_{-0.09}$ keV), consistent with \cite{Strickland2000}'s values within the range of the errors. 

As for the spectral fitting, \cite{Strickland2000} adopted a variable abundance, single temperature, \textsc{MEKAL} hot plasma model for the combined southern regions. Once they found best-fit metallicities, they adopted and froze these values when fitting the other regions while letting $kT$, $N^{\rm NGC253}_{\rm H}$, and normalization vary. Thus, by assuming a constant value for the metal abundances,  \cite{Strickland2000} did not constrain the metallicity gradients along the outflow as we did in this work. 

In \cite{Bauer2007}, X-ray spectra from the {\it XMM-Newton} Reflection Grating Spectrometer (RGS) were extracted along the starburst outflow. Four of their five regions overlap with ours, including two from the southern outflow, one from the center, and one from the northern outflow. They estimated the $kT$ of each region using line ratios of Si, Mg, Ne, and O. They did not measure other parameters (e.g., $N^{\rm NGC253}_{\rm H}$, or metal abundances) using spectral fits because of the limited statistics of the data. In the northern outflow, they found that the temperatures range from $0.21-0.61$~keV, in the center they range from $0.22-0.79$ keV, and in the southern regions they range from $0.21-0.46$ keV and $0.25-0.31$ keV. Our temperatures are about a factor of 1.4 larger in the first southern region and a factor of about 1.2 larger in all the other comparable regions. 

The discrepancies between \cite{Bauer2007} and this work may be a result of the different analysis performed. Rather than modeling the spectra of an entire region to find a best-fit temperature, \cite{Bauer2007} derived a temperature from the ratios of the same elements in different ionization states. They found that even within the same regions, the ratios yielded different temperatures. They noted that the different temperatures may be a result of the elements sampling different parts of the gas along the line of sight. If this is the case, then finding a best-fit temperature for the whole region would indeed result in a different value than temperatures from line ratios that sample different parts of the gas. Another physical explanation may be that there is a second temperature component in the southern regions. While we did not find a second temperature component to be required to improve fits there, it may still be physically present. If that is the case, then the second temperature component could be in better agreement with the results from \cite{Bauer2007}. 

 Another reason for the discrepancy between our work and Bauer may be the different instruments used. As mentioned before, \cite{Bauer2007} used X-ray spectra from the {\it XMM-Newton} Reflection Grating Spectrometer (RGS). This instrument only covers the $0.33-2.5$ keV energy range\footnote{https://www.cosmos.esa.int/web/xmm-newton/technical-details-rgs} in contrast to our {\it Chandra} data that is in the range of $0.5-7$ keV. Their lower temperatures may be a result of them probing the softer X-ray regime. Another explanation is that {\it Chandra} is currently less sensitive to softer X-rays than it was when it originally launched due to contamination in the detectors \citep{Plucinsky2003,Plucinsky2018}. About 60\% of our data was taken well after the contamination began affecting {\it Chandra's} soft X-ray sensitivity. As a result, our temperature values may be skewed upward.

\cite{Mitsuishi2013} used X-ray spectra from {\it XMM-Newton} and {\it Suzaku} to derive $kT$ values and metallicities from defined disk, superwind, and halo regions in NGC 253. Their superwind region was approximately the size and location of both of our southern regions, measuring 1.4 kpc $\times$ 0.9 kpc. They used an XSPEC model similar to ours, with two absorption components to account for the Milky Way's absorption and the intrinsic NGC~253 absorption. Different from our model, they opted for a two-temperature plasma (two \textsc{vapec} components) and incorporated a \textsc{zbremms} component to include contributions from point source emission, rather than a power-law component which is preferred for X-ray binaries \citep{Lehmer2013}. For their two temperatures, they found a warm component of $0.21$~keV and a warm-hot component of $0.62$~keV. These values are comparable to the range of our best-fit $kT$ values. Their metallicities are also comparable to ours in the southern hemisphere within the range of the errors. However, their $N^{\rm NGC253}_{\rm H}$ measurement is lower than ours with an upper limit of $0.05\times10^{22}\:\mathrm{cm^{-2}}$, while our value is $(0.15-0.18)\times10^{22}\:\mathrm{cm^{-2}}$.

\subsection{Comparison to M82 metal abundances and their distributions}\label{subsec:M82comp}
The analysis in this paper is similar to that performed on M82 in \cite{Lopez2020}. As the two nearest starburst galaxies where this analysis has been completed, it is worthwhile to compare their results for a more complete picture of the hot phase in galactic outflows. Differences in their metal enhancement trends, for example, would motivate investigations into how the differences arise and if they depend on galaxy properties. 

We detect the same metals in NGC~253 that were detected in M82, and we find that their distributions vary across the outflow. Ne is elevated in M82's northern hemisphere. However, for NGC~253, Ne has a flatter distribution with neither the northern nor southern outflow containing an enhancement. Mg and Fe have flat distributions in M82, while in NGC 253, they peak at the center and decrease with distance from the disk. Both M82 and NGC 253 have elevated amounts of S and Si in the central region and decrease with distance. Also, for both galaxies, the O abundance is not well constrained in regions of large column density because soft X-rays are attenuated. Both M82 and NGC~253 have an enhanced $\alpha/\rm{Fe}$ ratio in the outflow regions.

S and Si abundances peak in the center region of both galaxies and decrease with distance. However, in M82 the decrease in S and Si abundance is gradual, while in NGC 253 the decrease is sudden. As reported in \cite{Lopez2020}, the S and Si line fluxes originate predominantly from the hot component. In our analysis of NGC~253, only the central region has this hotter component, which is the likely explanation for why the S and Si peaks are so prominent in that location.   
\subsection{Connection to the CGM}

\cite{Das2019,Das2021} found $\alpha$-enhancement in their analysis of the Milky Way's CGM. \cite{Gupta2022} also found supersolar Ne/O in the Galactic eROSITA bubbles. The origin of the $\alpha$-enhancement or non-solar abundance ratio was unclear, but it was speculated that the enhancement may be a product of galactic outflows. 

In our analysis of the galactic wind of NGC~253, we find $\alpha$-enhancement already in the outflow regions. We also find super solar Ne/O ratios in the southern outflow regions. These elevated $\alpha/\rm{Fe}$ ratios and supersolar Ne/O in the wind may provide an origin for the observed metal enrichment of the Milky Way CGM. 

\subsection{Comparison to wind models}\label{sec:model_compare}

In order to provide a more detailed comparison to a theoretical galactic wind model, we take advantage of the smaller count threshold necessary to accurately constrain $kT$ and $N^{\rm NGC253}_{\rm H}$ (compared to the signal necessary to measure the metal abundances). Specifically, we create 42 regions along the outflow for spectral extraction that are 2.5\arcsec{} in height along the minor axis and 1\arcmin{} in width along the major axis. They span the same distance from the starburst as the larger regions we describe in Section~\ref{sec:methods}. We adopt an XSPEC model of \textsc{const*phabs*phabs*(vapec+powerlaw)} to fit the spectra associated with these regions. We freeze the metallicities in the \textsc{vapec} component to the best fit values found in the associated larger regions. Additionally, we look at lateral surface brightness profiles perpendicular to the wind axis for each region and find a $68\%$ and $95\%$ emission radius (as in Section~\ref{subsec:spectralresults}). From these radii, we calculate $n_{\rm e}$ for each region.

We first compare the best-fit $kT$ and $n_{\rm e}$ values to those expected in the \cite{CC85} (CC85) galactic wind model.
The CC85 model describes a SN-powered galactic wind that assumes spherical outflow geometry and adiabatic expansion. This model provides a good starting point for comparison because of its simplicity.

The controlling parameters of the CC85 model are the starburst radius $R$ and the total energy and mass-loading rates, $\dot{E}_\mathrm{T} = \alpha \times 3.1 \times 10^{41} \times (\dot{M}_\mathrm{SFR}/M_\odot \, \mathrm{yr^{-1}) \, [ergs \, s^{-1}]}$ and $\dot{M}_\mathrm{T} = \beta \times \dot{M}_\mathrm{SFR}$, respectively, where $\dot{M}_\mathrm{SFR}$ is the star-formation rate, and $\alpha$ and $\beta$ are dimensionless parameters. We take $R=200\,\mathrm{pc}$ \citep{Strickland2000,Bauer2007} and $\dot{M}_\mathrm{SFR}=2.5\,\mathrm{M_\odot \, yr^{-1}}$ \citep{Ott2005,Leroy2015,Bendo2015}. For the 68\%-ile contours, we find 
$\alpha_{68}  \simeq 0.32$ and $\beta_{68} \simeq 0.55$. For the 95\%-ile contours these parameters are $\alpha_{95} \simeq  0.21$ and $\beta_{95} \simeq  0.36$.

\begin{figure*}
    \centering
    \includegraphics[width=\textwidth]{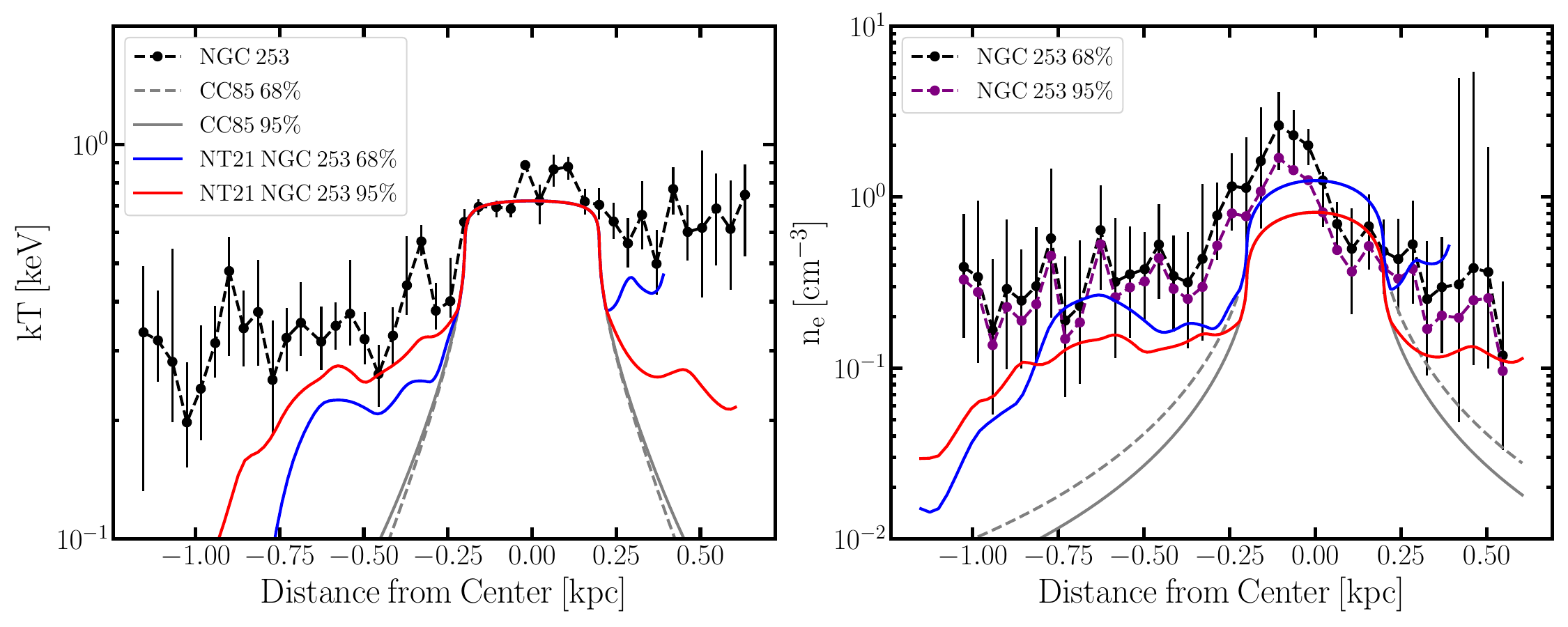}
    \caption{\textit{Left:} Best-fit $kT$ values (dashed line with `o' markers) as a function of distance along the minor axis from the center of the galaxy. The solid gray line shows the CC85 adiabatically expanding wind model $kT$ prediction for a 200 pc starburst radius and $95\%$ emission radius. The gray dashed line also shows the CC85 $kT$ predictions but for a $68\%$ emission radius. The red line shows $kT$ predictions from a non-spherical wind model using the $95\%$ emission radius. The blue line is the same as the red but for $68\%$ emission radius. \textit{Right:} Same models as the right except now $n_e$ values are plotted for a $95\%$ (black dashed line with `o' markers) and for a $68\%$ (purple dashed line with `o' markers) emission radius. We find better agreement with the non-spherical models and find the CC85 model does not characterize the $kT$ and $n_e$ profiles well.}
    \label{fig:42reg_model}
\end{figure*}

In Figure~\ref{fig:42reg_model}, we plot the observed $kT$ and $n_{\rm e}$ values (black   and purple dashed lines with `o' markers) with that of the spherical adiabatic CC85 model (gray lines). The disagreement between the CC85 model with that of the best-fit, observed values primarily widens when the flow leaves the wind-driving region (i.e., when $r>R$). We see that the adiabatic spherical model produces temperature and density profiles that fall off faster than that of the observed values.

This may be a result of missing processes like non-spherical areal divergence \citep{Nguyen2021}. In NGC~253, the super star clusters within the nuclear core are likely arranged in a ring-like geometry \citep{Levy2022}. \citet{Nguyen2022} recently showed that energy and mass injection in a ring-like geometry naturally produces a bi-conical outflow, where the hot phase is cylindrically collimated. When we include the effect of non-spherical flow geometry into the wind model (red and blue lines in Figure~\ref{fig:42reg_model}), we find better agreement with the profiles. 

\begin{figure*}
    \centering
    \includegraphics[width=\textwidth]{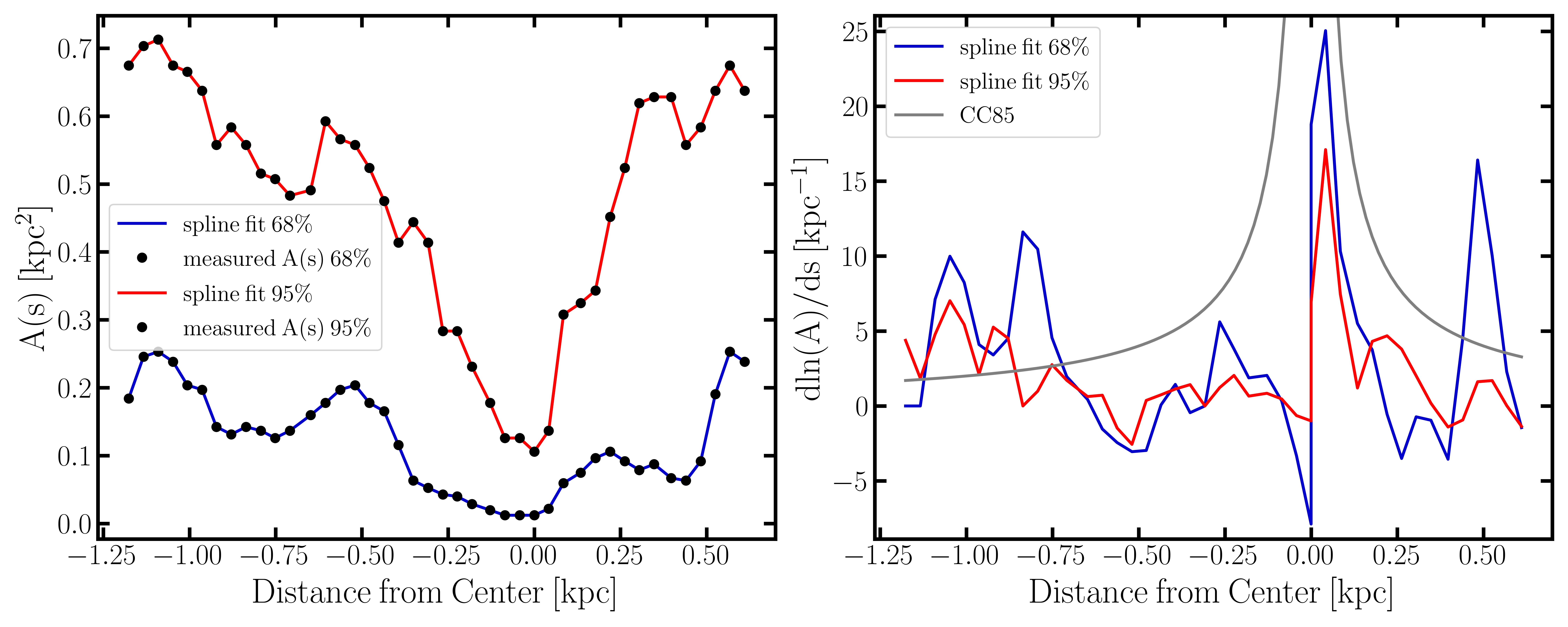}
    \caption{\textit{Left:} The non-spherical area function for NGC 253’s hot wind derived from the observed X-ray (0.5-7 keV) surface brightness contours perpendicular to the minor axis as a function of height s above and below the starburst disk at s = 0. The black circles and diamonds show the lateral size of the contour enclosing 68 \% and 95\% of the emission, respectively, as a function of distance along the minor axis. The blue and red lines are a cubic spline, fit to both the southern and northern sides independently, for the 68\% and 95\% contours, respectively. \textit{Right:} The areal expansion rate for the cubic spline fits to A(s) and for spherical expansion (gray line). These profiles demonstrate that the wind does not have the geometry assumed by the CC85 model.}
    \label{fig:areas}
\end{figure*}

The enclosed $68\%$-ile and $95\%$-ile volumes, which were used to determine $n_{\rm e}$, can also be used to define an outflow geometry $A(s)= \pi r(s) ^2$, where $r(s)$ is the radius of the $68\%$-ile or $95\%$-ile volume at height $s$ perpendicular to the wind axis. In Figure \ref{fig:areas}, we plot $A(s)$ and $d\ln A/ds$, where the latter term describes the areal expansion rate. In the left panel of Figure \ref{fig:areas}, we see that the derived $A(s)$ is highly non-spherical. In the right panel, we see that the derived $A(s)$ (red and blue lines) undergoes an areal expansion rate slower than that of spherical (gray line). These profiles demonstrate that the wind does not have the geometry assumed by the CC85 model.

\begin{figure*}
    \centering
    \includegraphics[width=\textwidth]{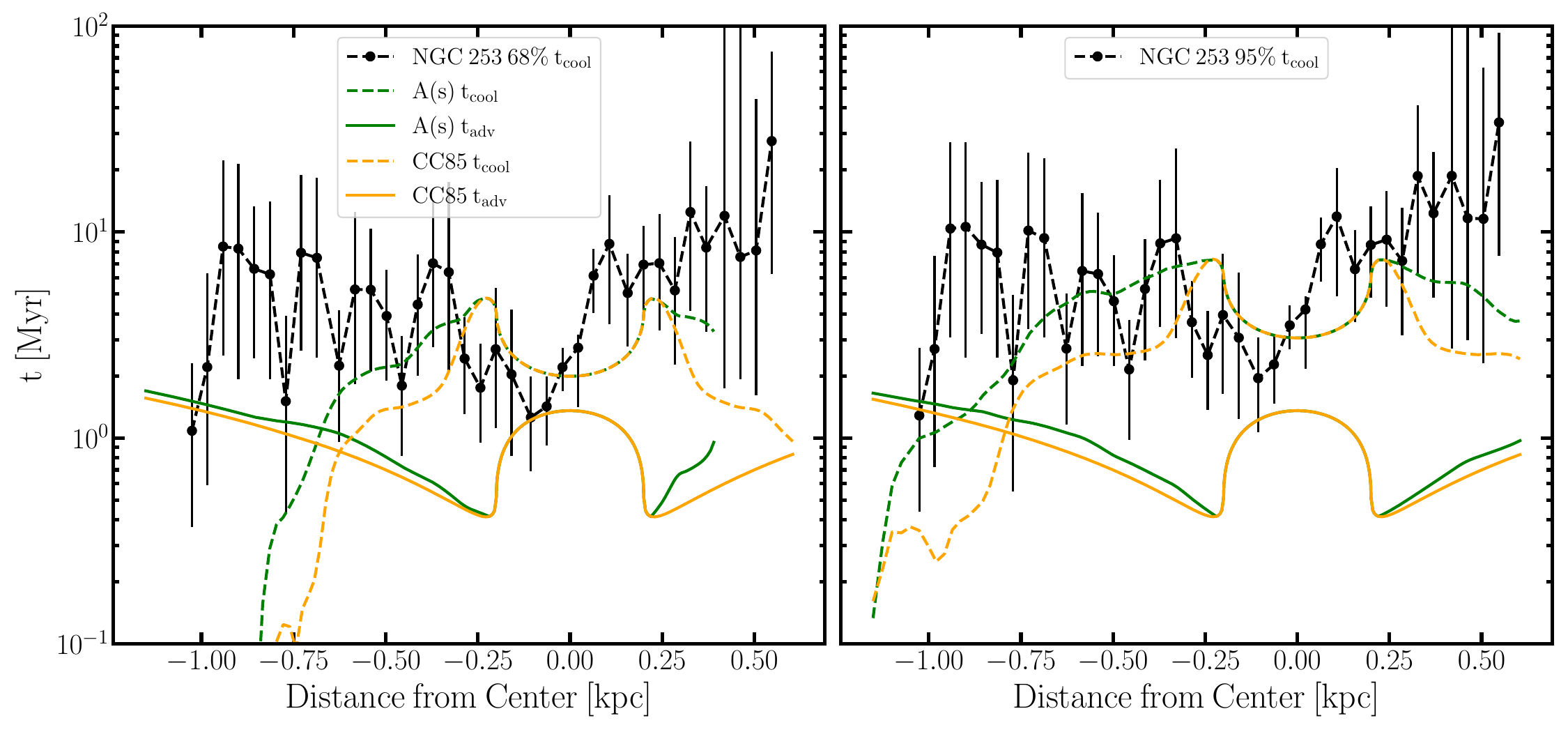}
    \caption{\textit{Left:} Calculated cooling times for a 68\% (black dashed line with `o' markers). The solid orange line is the CC85 predicted advection time and the dashed orange line is the CC85 cooling time. The solid green line is the $A(s)$ predicted advection time and the dashed orange line is the $A(s)$ cooling time. \textit{Right:} The same as the left but with the values for the 95\% emission. We find that the cooling times for the 68\% emission are of order the prediction advection times. This may indicate the presence of bulk radiative cooling in the outflow.}
    \label{fig:cooling}
\end{figure*}

In Figure~\ref{fig:cooling} we show the calculated cooling times ($t_{\rm cool}$) for a $68\%$ and $95\%$ emission radius, along with the  advection times ($t_{adv}$) from both the spherical (orange lines) and non-spherical (green lines) wind models described above. \edit1{The advection times are calculated as $t_{\rm adv} = r/v$, where $r$ is the distance along the minor axis to the central starburst and $v$ is the velocity of the wind as it advects, predicted from the models.} As noted in Section~\ref{sec:results}, we assume a filling factor of $f = 1$, therefore the cooling times derived are upper limits. Using the 68\% emission results, we find that the measured $t_{\rm cool}$ values are of order (but somewhat larger than) the predicted advection times from both the CC85 and $A(s)$ models. Similarly, using the derived values of the thermalization efficiency ($\alpha$) and mass-loading parameter ($\beta$) in the core (using either the 68\% or 95\% contours), we estimate an advection timescale of order $\simeq0.5-1$\,Myr on $\simeq0.2-0.5$\,kpc scales along the wind axis.  

As shown by \cite{Thompson2016}, when the cooling time is less than a few times longer then the advection time, strong bulk radiative cooling may result. Thus, on the basis of the comparison in Figure~\ref{fig:cooling}, it is possible that the hot outflow from NGC 253 goes through bulk radiative cooling, perhaps changing the character of both the X-ray emission and the H$\alpha$ emissions along the wind axis. Bulk cooling from the hot phase is important because it provides a physical origin for high velocity cool phase outflows \citep{Wang1995,Silich2004}.

\subsection{Mass Outflow Rates}\label{sec:mass_outflow}

We calculate an estimate of the outflow mass as $M = n_e m_H V/ 1.2f^{1/2}$ where the factor of 1.2 is from our assumed relation of $n_e=1.2n_{\rm H}$, $f$ is the filling factor which we assume 1, and $V$ is the cylindrical volume using a $95\%$ emission radius. We also calculate a mass outflow rate as $\dot M = n_e m_H r v/ 1.2f^{1/2}$ where $r$ is the vertical distance traveled, and $v$ is the outflow velocity. For $r$ we use 0.43 kpc in the northern and 0.95 kpc in the southern outflow, where we ignore the 200 pc central starburst region in order to solely capture the outflowing gas. For $v$ we assume a velocity of 1000 $\rm{km/s}$ based on the models described in Section~\ref{sec:model_compare}. We find a mass in the northern outflow of $1.2\times10^{6}\:M_{\odot}$ and a mass of $3.0\times10^{6}\:M_{\odot}$ in the southern outflow. For the mass outflow rates with the assumptions from above, we find a rate of $2.8\:M_{\odot}/\rm{yr}$ in the northern outflow and $3.2\:M_{\odot}/\rm{yr}$ in the southern outflow. 


We compare our mass outflow rate estimates with those from \cite{Strickland2000} and find that they are similar. \cite{Strickland2000} estimated a mass outflow rate upper limit of $2.2v_3\:M_{\odot}/\rm{yr}$ where $v_3$ is the outflow velocity in units of $3000\:\rm{km/s}$. Although the mass outflow rates are similar, the assumptions used in the estimates are different and warrant a discussion.


\cite{Strickland2000} used a velocity of 3000 km/s based on standard mass and energy injection rates from \cite{Leitherer1995}. However, we find in our models described in Section~\ref{sec:model_compare}, that the velocity profile produces much lower values. Based on our derived CC85 parameters of $\alpha_{95} \simeq  0.21$ and $\beta_{95} \simeq  0.36$, we find that the velocity of the wind is closer to 1000 km/s. \cite{Strickland2000} also used a filling factor of 0.4. They asserted than on scales of $\lesssim$ a few hundred parsecs, the pressures of the X-ray and H$\alpha$ emitting phase are similar. By equating their derived pressure (which they write in terms of the filling factor) to the H$\alpha$ pressure from \cite{McCarthy1987}, they were able to derive a filling factor of 0.4.

If we use the \cite{Strickland2000} assumptions with our formulae, we find that the derived mass outflow rates are too large when compared to the cooler phases. Assuming $f=0.4$ and $v=3000$ km~s$^{-1}$, we find a hot phase mass outflow rate of $13.5\:M_{\odot}/\rm{yr}$ in the northern outflow and $15.3\:M_{\odot}/\rm{yr}$ in the southern outflow. In the cool molecular phase, the estimate for the mass outflow rate is $14-39\:M_{\odot}/\rm{yr}$ \citep{Krieger2019}, while in the warm ionized phase, the mass outflow rate is $\approx4\:M_{\odot}/\rm{yr}$ \citep{Westmoquette2011,Krieger2019}. The \cite{Strickland2000} assumptions would result in a hot phase outflow rate on par with the molecular phase rate and greater than the warm phase. The results from the \cite{Strickland2000} assumptions lead to mass outflow rates dominated by the hot phase.

\subsection{Charge Exchange contribution}\label{sec:CX}

In order to acquire the most accurate measurements of outflow properties, the spectral model needs to include all the components that contribute significantly to the X-ray emission. Previous work has shown that charge exchange (CX), the stripping of an electron from a neutral atom by an ion, is one of these components. In particular, \cite{Lopez2020} found that when including CX in spectral models of the outflows of M82, CX contributes a significant portion (up to 25\%) of the broad-band ($0.5-7$~keV) X-ray emission. Similarly, for several nearby star-forming galaxies (including NCC~253), \cite{Liu2012} showed that the K$\alpha$ triplet of He-like ions required a CX component to account for the observed line ratios. In the context of galactic winds, CX represents the emission of the hot phase interacting with the cooler phases where the hot ionized gas strips electrons from the cooler neutral gas. 

\begin{figure}
    \centering
    \includegraphics[width=0.47\textwidth]{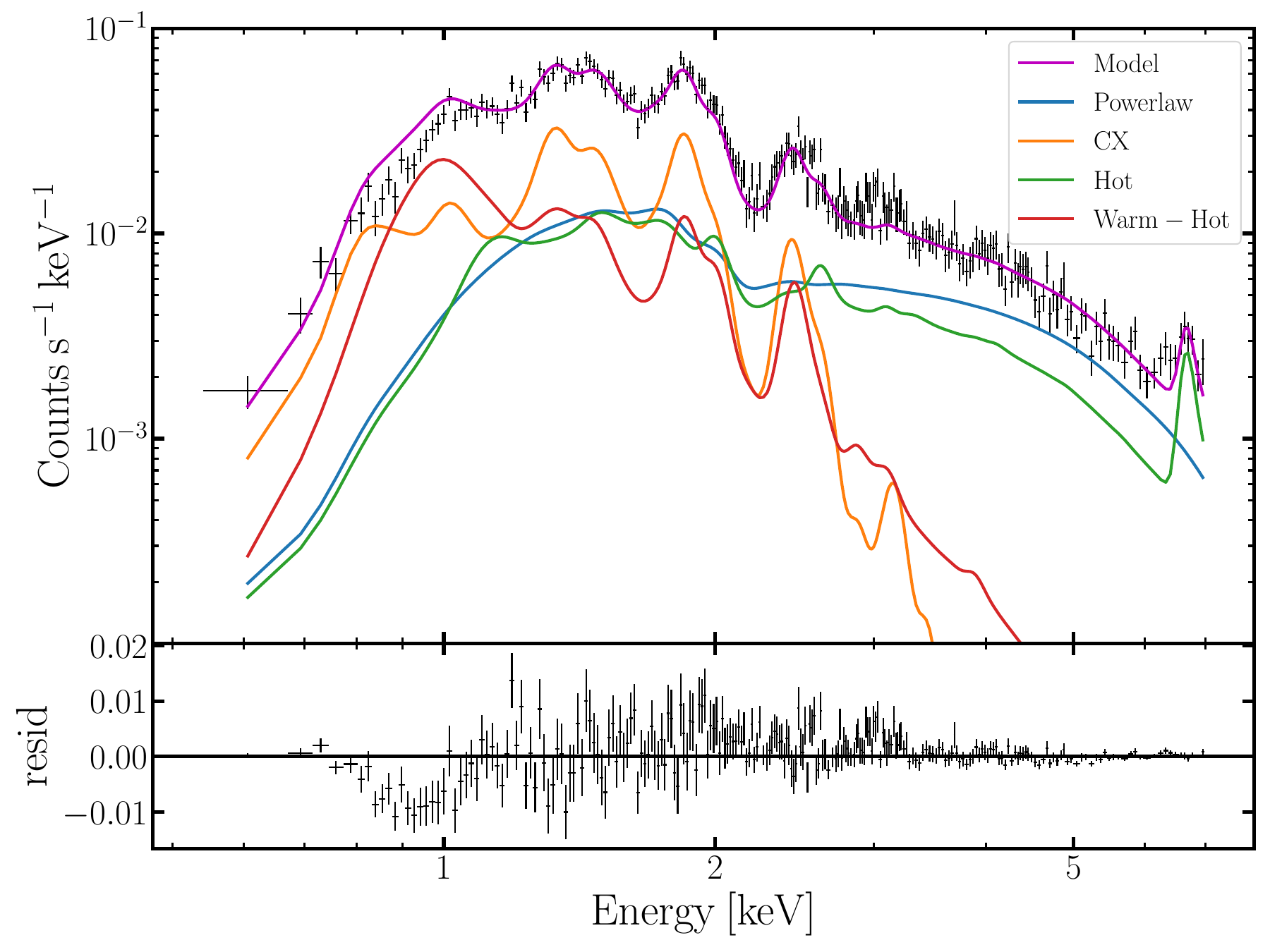}
     \caption{Spectra of the central region for Observation 20343 with the total spectral model, individual model components, and residuals plotted. The CX component accounts for significant fraction of spectra.}
    \label{fig:cx_spectra}
\end{figure}

In our analysis, we find that three regions' spectral fits were statistically significantly better, based on F-tests, with the inclusion of a CX component: the center and the two southern regions. At these locations, CX accounts for a significant fraction of the emission. As an example, in Figure~\ref{fig:cx_spectra} we show the spectra for Observation 20343 in the central region and note that CX makes up a significant fraction of the spectra. In the center region, the CX contribution is 42$\%$. For region S1, it is 20$\%$, and for region S2, it is 33$\%$ of the total broad-band X-ray emission. We note that the lack of a CX contribution in the northern outflow may be because of the high column density there, obscuring the soft X-rays necessary to distinguish the CX emission. 

\section{Conclusions} \label{sec:con}

In this paper, we analyze $\approx$365~ks of archival {\it Chandra} data from NGC 253 to produce constraints on its starburst driven outflow that spans 1.1~kpc south and to 0.6~kpc north from the galaxy's center. We extract spectra and use models to find the best-fit parameters ($kT$, $N^{\rm NGC253}_{\rm H}$, and metal abundances) in five regions along the outflow (Figure~\ref{fig:ngc253_box}) and in nine regions of the galaxy's disk (Figure~\ref{fig:disk_reg}). 

With the exception of Ne which has a nearly flat distribution, the metal abundances peak in the galactic disk and decrease sharply along the minor axis (Figure~\ref{fig:abund} and Table~\ref{table:fitresults}). The distributions indicate significant dilution outside of the starburst region. For $kT$ and $N^{\rm NGC253}_{\rm H}$, the values also peak in the galactic disk and decrease along the minor axis. The decrease in abundances along the outflow may suggest mass loading or indicate other physics should be considered, such as non-equilibrium ionization.

We constrain the extent of the outflow in the regions and calculate $n_e$, pressure, and cooling times (Table~\ref{table:ne}). With these measurements we provide estimates of the hot phase mass outflow rates in Section~\ref{sec:mass_outflow}, finding them to be $2.8\:M_{\odot}/\rm{yr}$ in the northern outflow and $3.2\:M_{\odot}/\rm{yr}$ in the southern outflow. We also find that the measured cooling times are of order the advection times predicted from models potentially indicating the presence of bulk radiative cooling in the outflow in Section~\ref{sec:model_compare}. The models also show that incorporating non-spherical geometry provides a better match to observed $kT$ and $n_e$ profiles. We also find that the central region and southern outflow have significant contribution of $20-42$\% to the broad-band X-ray emission from a CX component in Section~\ref{sec:CX}.

In the future, expanding this approach to other starburst galaxies will reveal whether outflow properties vary with host galaxy properties, such as stellar mass. \cite{Chisholm2018} already showed that there is an inverse relationship between the metal loading factor and galaxy stellar mass in warm gas outflows. It would be worth investigating whether this relationship also holds in the hot phase, where most of the metals are being driven out.

Additionally, understanding of the hot gas in starburst-driven outflows will advance tremendously with the launch of XRISM~\citep{XRISM}. The mission will have higher spectral resolution, facilitating measurement of the hot wind velocity and enabling better constraints on the wind energy \citep{Hodges-Kluck} and improving our hot phase mass outflow rate estimates. 

\acknowledgements
We thank Ann Hornschemeier for providing us with the NGC~253 \textit{Chandra} data and allowing us to use it for this work. We also thank Adam Leroy and the OSU Galaxy/ISM Meeting for useful discussions. SL and LAL were supported by NASA's Astrophysics Data Analysis Program under grant number 80NSSC22K0496. LAL also acknowledges support by the Simons Foundation, the Heising-Simons Foundation, and a Cottrell Scholar Award from the Research Corporation for Science Advancement. ADB acknowledges support from NSF-AST2108140. This work was performed in part at the Simons Foundation Flatiron Institute's Center for Computational Astrophysics during LAL's tenure as an IDEA Scholar. SM gratefully acknowledges the support provided by the National Aeronautics and Space Administration through Chandra Award Number AR2-23014X issued by the Chandra X-ray Center, which is operated by the Smithsonian Astrophysical Observatory for and on behalf of the National Aeronautics Space Administration under contract NAS8-03060. DDN and TAT are supported by NSF grant 1516967, NASA ATP 80NSSC18K0526, and NASA 21-ASTRO21-0174. AS is supported by an NSF Astronomy and Astrophysics Postdoctoral Fellowship under award AST-1903834.

\software{CIAO (v4.14; \citealt{CIAO2006}), XSPEC (v12.12.1; \citealt{XSPEC})}

\clearpage

\bibliographystyle{aasjournal}
\bibliography{sample63}{}

\begin{thebibliography}{}
\expandafter\ifx\csname natexlab\endcsname\relax\def\natexlab#1{#1}\fi
\providecommand{\url}[1]{\href{#1}{#1}}
\providecommand{\dodoi}[1]{doi:~\href{http://doi.org/#1}{\nolinkurl{#1}}}
\providecommand{\doeprint}[1]{\href{http://ascl.net/#1}{\nolinkurl{http://ascl.net/#1}}}
\providecommand{\doarXiv}[1]{\href{https://arxiv.org/abs/#1}{\nolinkurl{https://arxiv.org/abs/#1}}}

\bibitem[{{Arnaud}(1996)}]{XSPEC}
{Arnaud}, K.~A. 1996, in Astronomical Society of the Pacific Conference Series,
  Vol. 101, Astronomical Data Analysis Software and Systems V, ed. G.~H.
  {Jacoby} \& J.~{Barnes}, 17

\bibitem[{{Asplund} {et~al.}(2009){Asplund}, {Grevesse}, {Sauval}, \&
  {Scott}}]{Asplund2009}
{Asplund}, M., {Grevesse}, N., {Sauval}, A.~J., \& {Scott}, P. 2009, \araa, 47,
  481, \dodoi{10.1146/annurev.astro.46.060407.145222}

\bibitem[{{Bauer} {et~al.}(2007){Bauer}, {Pietsch}, {Trinchieri},
  {Breitschwerdt}, {Ehle}, \& {Read}}]{Bauer2007}
{Bauer}, M., {Pietsch}, W., {Trinchieri}, G., {et~al.} 2007, \aap, 467, 979,
  \dodoi{10.1051/0004-6361:20066340}

\bibitem[{{Beck} {et~al.}(2022){Beck}, {Lebouteiller}, {Madden}, {Iserlohe},
  {Krabbe}, {Ramambason}, {Fischer}, {Ka{\'z}mierczak-Barthel}, {Latzko}, \&
  {P{\'e}rez-Beaupuits}}]{Beck2022}
{Beck}, A., {Lebouteiller}, V., {Madden}, S.~C., {et~al.} 2022, \aap, 665, A85,
  \dodoi{10.1051/0004-6361/202243822}

\bibitem[{{Bendo} {et~al.}(2015){Bendo}, {Beswick}, {D'Cruze}, {Dickinson},
  {Fuller}, \& {Muxlow}}]{Bendo2015}
{Bendo}, G.~J., {Beswick}, R.~J., {D'Cruze}, M.~J., {et~al.} 2015, \mnras, 450,
  L80, \dodoi{10.1093/mnrasl/slv053}

\bibitem[{{Bolatto} {et~al.}(2013){Bolatto}, {Warren}, {Leroy}, {Walter},
  {Veilleux}, {Ostriker}, {Ott}, {Zwaan}, {Fisher}, {Weiss}, {Rosolowsky}, \&
  {Hodge}}]{Bolatto2013}
{Bolatto}, A.~D., {Warren}, S.~R., {Leroy}, A.~K., {et~al.} 2013, \nat, 499,
  450, \dodoi{10.1038/nature12351}

\bibitem[{{Chevalier} \& {Clegg}(1985)}]{CC85}
{Chevalier}, R.~A., \& {Clegg}, A.~W. 1985, \nat, 317, 44,
  \dodoi{10.1038/317044a0}

\bibitem[{{Chisholm} {et~al.}(2018){Chisholm}, {Tremonti}, \&
  {Leitherer}}]{Chisholm2018}
{Chisholm}, J., {Tremonti}, C., \& {Leitherer}, C. 2018, \mnras, 481, 1690,
  \dodoi{10.1093/mnras/sty2380}

\bibitem[{{Das} {et~al.}(2021){Das}, {Mathur}, {Gupta}, \&
  {Krongold}}]{Das2021}
{Das}, S., {Mathur}, S., {Gupta}, A., \& {Krongold}, Y. 2021, \apj, 918, 83,
  \dodoi{10.3847/1538-4357/ac0e8e}

\bibitem[{{Das} {et~al.}(2019){Das}, {Mathur}, {Nicastro}, \&
  {Krongold}}]{Das2019}
{Das}, S., {Mathur}, S., {Nicastro}, F., \& {Krongold}, Y. 2019, \apjl, 882,
  L23, \dodoi{10.3847/2041-8213/ab3b09}

\bibitem[{{de Vaucouleurs} {et~al.}(1991){de Vaucouleurs}, {de Vaucouleurs},
  {Corwin}, {Buta}, {Paturel}, \& {Fouque}}]{RC3}
{de Vaucouleurs}, G., {de Vaucouleurs}, A., {Corwin}, Herold~G., J., {et~al.}
  1991, {Third Reference Catalogue of Bright Galaxies}

\bibitem[{{Dere} {et~al.}(1997){Dere}, {Landi}, {Mason}, {Monsignori Fossi}, \&
  {Young}}]{CHIANTI}
{Dere}, K.~P., {Landi}, E., {Mason}, H.~E., {Monsignori Fossi}, B.~C., \&
  {Young}, P.~R. 1997, \aaps, 125, 149, \dodoi{10.1051/aas:1997368}

\bibitem[{{Dickey} \& {Lockman}(1990)}]{DL1990}
{Dickey}, J.~M., \& {Lockman}, F.~J. 1990, \araa, 28, 215,
  \dodoi{10.1146/annurev.aa.28.090190.001243}

\bibitem[{{Fielding} {et~al.}(2017){Fielding}, {Quataert}, {McCourt}, \&
  {Thompson}}]{Fielding2017}
{Fielding}, D., {Quataert}, E., {McCourt}, M., \& {Thompson}, T.~A. 2017,
  \mnras, 466, 3810, \dodoi{10.1093/mnras/stw3326}

\bibitem[{{Finlator} \& {Dav{\'e}}(2008)}]{Finlator2008}
{Finlator}, K., \& {Dav{\'e}}, R. 2008, \mnras, 385, 2181,
  \dodoi{10.1111/j.1365-2966.2008.12991.x}

\bibitem[{{Fruscione} {et~al.}(2006){Fruscione}, {McDowell}, {Allen},
  {Brickhouse}, {Burke}, {Davis}, {Durham}, {Elvis}, {Galle}, {Harris},
  {Huenemoerder}, {Houck}, {Ishibashi}, {Karovska}, {Nicastro}, {Noble},
  {Nowak}, {Primini}, {Siemiginowska}, {Smith}, \& {Wise}}]{CIAO2006}
{Fruscione}, A., {McDowell}, J.~C., {Allen}, G.~E., {et~al.} 2006, in Society
  of Photo-Optical Instrumentation Engineers (SPIE) Conference Series, Vol.
  6270, Society of Photo-Optical Instrumentation Engineers (SPIE) Conference
  Series, ed. D.~R. {Silva} \& R.~E. {Doxsey}, 62701V,
  \dodoi{10.1117/12.671760}

\bibitem[{{Gupta} {et~al.}(2022){Gupta}, {Mathur}, {Kingsbury}, {Das}, \&
  {Krongold}}]{Gupta2022}
{Gupta}, A., {Mathur}, S., {Kingsbury}, J., {Das}, S., \& {Krongold}, Y. 2022,
  arXiv e-prints, arXiv:2201.09915.
\newblock \doarXiv{2201.09915}

\bibitem[{{Hodges-Kluck} {et~al.}(2019){Hodges-Kluck}, {Lopez}, {Yukita},
  {Ptak}, {Swartz}, {Tzanavaris}, {Veilleux}, \& {Bregman}}]{Hodges-Kluck}
{Hodges-Kluck}, E., {Lopez}, L.~A., {Yukita}, M., {et~al.} 2019, \baas, 51,
  257.
\newblock \doarXiv{1903.09692}

\bibitem[{{Krieger} {et~al.}(2019){Krieger}, {Bolatto}, {Walter}, {Leroy},
  {Zschaechner}, {Meier}, {Ott}, {Weiss}, {Mills}, {Levy}, {Veilleux}, \&
  {Gorski}}]{Krieger2019}
{Krieger}, N., {Bolatto}, A.~D., {Walter}, F., {et~al.} 2019, \apj, 881, 43,
  \dodoi{10.3847/1538-4357/ab2d9c}

\bibitem[{{Lehmer} {et~al.}(2013){Lehmer}, {Wik}, {Hornschemeier}, {Ptak},
  {Antoniou}, {Argo}, {Bechtol}, {Boggs}, {Christensen}, {Craig}, {Hailey},
  {Harrison}, {Krivonos}, {Leyder}, {Maccarone}, {Stern}, {Venters}, {Zezas},
  \& {Zhang}}]{Lehmer2013}
{Lehmer}, B.~D., {Wik}, D.~R., {Hornschemeier}, A.~E., {et~al.} 2013, \apj,
  771, 134, \dodoi{10.1088/0004-637X/771/2/134}

\bibitem[{{Lehnert} \& {Heckman}(1995)}]{LH1995}
{Lehnert}, M.~D., \& {Heckman}, T.~M. 1995, \apjs, 97, 89,
  \dodoi{10.1086/192137}

\bibitem[{{Leitherer} \& {Heckman}(1995)}]{Leitherer1995}
{Leitherer}, C., \& {Heckman}, T.~M. 1995, \apjs, 96, 9, \dodoi{10.1086/192112}

\bibitem[{{Leroy} {et~al.}(2015){Leroy}, {Bolatto}, {Ostriker}, {Rosolowsky},
  {Walter}, {Warren}, {Donovan Meyer}, {Hodge}, {Meier}, {Ott}, {Sandstrom},
  {Schruba}, {Veilleux}, \& {Zwaan}}]{Leroy2015}
{Leroy}, A.~K., {Bolatto}, A.~D., {Ostriker}, E.~C., {et~al.} 2015, \apj, 801,
  25, \dodoi{10.1088/0004-637X/801/1/25}

\bibitem[{{Leroy} {et~al.}(2021){Leroy}, {Schinnerer}, {Hughes}, {Rosolowsky},
  {Pety}, {Schruba}, {Usero}, {Blanc}, {Chevance}, {Emsellem}, {Faesi},
  {Herrera}, {Liu}, {Meidt}, {Querejeta}, {Saito}, {Sandstrom}, {Sun},
  {Williams}, {Anand}, {Barnes}, {Behrens}, {Belfiore}, {Benincasa},
  {Be{\v{s}}li{\'c}}, {Bigiel}, {Bolatto}, {den Brok}, {Cao}, {Chandar},
  {Chastenet}, {Chiang}, {Congiu}, {Dale}, {Deger}, {Eibensteiner}, {Egorov},
  {Garc{\'\i}a-Rodr{\'\i}guez}, {Glover}, {Grasha}, {Henshaw}, {Ho}, {Kepley},
  {Kim}, {Klessen}, {Kreckel}, {Koch}, {Kruijssen}, {Larson}, {Lee}, {Lopez},
  {Machado}, {Mayker}, {McElroy}, {Murphy}, {Ostriker}, {Pan}, {Pessa},
  {Puschnig}, {Razza}, {S{\'a}nchez-Bl{\'a}zquez}, {Santoro}, {Sardone},
  {Scheuermann}, {Sliwa}, {Sormani}, {Stuber}, {Thilker}, {Turner}, {Utomo},
  {Watkins}, \& {Whitmore}}]{Leroy2021}
{Leroy}, A.~K., {Schinnerer}, E., {Hughes}, A., {et~al.} 2021, \apjs, 257, 43,
  \dodoi{10.3847/1538-4365/ac17f3}

\bibitem[{{Levy} {et~al.}(2022){Levy}, {Bolatto}, {Leroy}, {Sormani}, {Emig},
  {Gorski}, {Lenki{\'c}}, {Mills}, {Tarantino}, {Teuben}, {Veilleux}, \&
  {Walter}}]{Levy2022}
{Levy}, R.~C., {Bolatto}, A.~D., {Leroy}, A.~K., {et~al.} 2022, arXiv e-prints,
  arXiv:2206.04700.
\newblock \doarXiv{2206.04700}

\bibitem[{{Liu} {et~al.}(2012){Liu}, {Wang}, \& {Mao}}]{Liu2012}
{Liu}, J., {Wang}, Q.~D., \& {Mao}, S. 2012, \mnras, 420, 3389,
  \dodoi{10.1111/j.1365-2966.2011.20263.x}

\bibitem[{{Lopez} {et~al.}(2020){Lopez}, {Mathur}, {Nguyen}, {Thompson}, \&
  {Olivier}}]{Lopez2020}
{Lopez}, L.~A., {Mathur}, S., {Nguyen}, D.~D., {Thompson}, T.~A., \& {Olivier},
  G.~M. 2020, \apj, 904, 152, \dodoi{10.3847/1538-4357/abc010}

\bibitem[{{McCarthy} {et~al.}(1987){McCarthy}, {van Breugel}, \&
  {Heckman}}]{McCarthy1987}
{McCarthy}, P.~J., {van Breugel}, W., \& {Heckman}, T. 1987, \aj, 93, 264,
  \dodoi{10.1086/114309}

\bibitem[{{McCormick} {et~al.}(2013){McCormick}, {Veilleux}, \&
  {Rupke}}]{McCormick2013}
{McCormick}, A., {Veilleux}, S., \& {Rupke}, D. S.~N. 2013, \apj, 774, 126,
  \dodoi{10.1088/0004-637X/774/2/126}

\bibitem[{{Mills} {et~al.}(2021){Mills}, {Gorski}, {Emig}, {Bolatto}, {Levy},
  {Leroy}, {Ginsburg}, {Henshaw}, {Zschaechner}, {Veilleux}, {Tanaka}, {Meier},
  {Walter}, {Krieger}, \& {Ott}}]{Mills2021}
{Mills}, E.~A.~C., {Gorski}, M., {Emig}, K.~L., {et~al.} 2021, \apj, 919, 105,
  \dodoi{10.3847/1538-4357/ac0fe8}

\bibitem[{{Mitsuishi} {et~al.}(2013){Mitsuishi}, {Yamasaki}, \&
  {Takei}}]{Mitsuishi2013}
{Mitsuishi}, I., {Yamasaki}, N.~Y., \& {Takei}, Y. 2013, \pasj, 65, 44,
  \dodoi{10.1093/pasj/65.2.44}

\bibitem[{{M{\"u}ller-S{\'a}nchez} {et~al.}(2010){M{\"u}ller-S{\'a}nchez},
  {Gonz{\'a}lez-Mart{\'\i}n}, {Fern{\'a}ndez-Ontiveros}, {Acosta-Pulido}, \&
  {Prieto}}]{Muller-Sanchez2010}
{M{\"u}ller-S{\'a}nchez}, F., {Gonz{\'a}lez-Mart{\'\i}n}, O.,
  {Fern{\'a}ndez-Ontiveros}, J.~A., {Acosta-Pulido}, J.~A., \& {Prieto}, M.~A.
  2010, \apj, 716, 1166, \dodoi{10.1088/0004-637X/716/2/1166}

\bibitem[{{Nguyen} \& {Thompson}(2021)}]{Nguyen2021}
{Nguyen}, D.~D., \& {Thompson}, T.~A. 2021, \mnras, 508, 5310,
  \dodoi{10.1093/mnras/stab2910}

\bibitem[{{Nguyen} \& {Thompson}(2022)}]{Nguyen2022}
---. 2022, \apjl, 935, L24, \dodoi{10.3847/2041-8213/ac86c3}

\bibitem[{{Oppenheimer} \& {Dav{\'e}}(2008)}]{Opp2008}
{Oppenheimer}, B.~D., \& {Dav{\'e}}, R. 2008, \mnras, 387, 577,
  \dodoi{10.1111/j.1365-2966.2008.13280.x}

\bibitem[{{Ott} {et~al.}(2005){Ott}, {Weiss}, {Henkel}, \& {Walter}}]{Ott2005}
{Ott}, J., {Weiss}, A., {Henkel}, C., \& {Walter}, F. 2005, \apj, 629, 767,
  \dodoi{10.1086/431661}

\bibitem[{{Peeples} \& {Shankar}(2011)}]{Peeples2011}
{Peeples}, M.~S., \& {Shankar}, F. 2011, \mnras, 417, 2962,
  \dodoi{10.1111/j.1365-2966.2011.19456.x}

\bibitem[{{Plucinsky} {et~al.}(2018){Plucinsky}, {Bogdan}, {Marshall}, \&
  {Tice}}]{Plucinsky2018}
{Plucinsky}, P.~P., {Bogdan}, A., {Marshall}, H.~L., \& {Tice}, N.~W. 2018, in
  Society of Photo-Optical Instrumentation Engineers (SPIE) Conference Series,
  Vol. 10699, Space Telescopes and Instrumentation 2018: Ultraviolet to Gamma
  Ray, ed. J.-W.~A. {den Herder}, S.~{Nikzad}, \& K.~{Nakazawa}, 106996B,
  \dodoi{10.1117/12.2312748}

\bibitem[{{Plucinsky} {et~al.}(2003){Plucinsky}, {Schulz}, {Marshall}, {Grant},
  {Chartas}, {Sanwal}, {Teter}, {Vikhlinin}, {Edgar}, {Wise}, {Allen},
  {Virani}, {DePasquale}, \& {Raley}}]{Plucinsky2003}
{Plucinsky}, P.~P., {Schulz}, N.~S., {Marshall}, H.~L., {et~al.} 2003, in
  Society of Photo-Optical Instrumentation Engineers (SPIE) Conference Series,
  Vol. 4851, X-Ray and Gamma-Ray Telescopes and Instruments for Astronomy., ed.
  J.~E. {Truemper} \& H.~D. {Tananbaum}, 89--100, \dodoi{10.1117/12.461473}

\bibitem[{{Rekola} {et~al.}(2005){Rekola}, {Richer}, {McCall}, {Valtonen},
  {Kotilainen}, \& {Flynn}}]{Rekola2005}
{Rekola}, R., {Richer}, M.~G., {McCall}, M.~L., {et~al.} 2005, \mnras, 361,
  330, \dodoi{10.1111/j.1365-2966.2005.09166.x}

\bibitem[{{Rubin} {et~al.}(2014){Rubin}, {Prochaska}, {Koo}, {Phillips},
  {Martin}, \& {Winstrom}}]{Rubin2014}
{Rubin}, K. H.~R., {Prochaska}, J.~X., {Koo}, D.~C., {et~al.} 2014, \apj, 794,
  156, \dodoi{10.1088/0004-637X/794/2/156}

\bibitem[{{Silich} {et~al.}(2004){Silich}, {Tenorio-Tagle}, \&
  {Rodr{\'\i}guez-Gonz{\'a}lez}}]{Silich2004}
{Silich}, S., {Tenorio-Tagle}, G., \& {Rodr{\'\i}guez-Gonz{\'a}lez}, A. 2004,
  \apj, 610, 226, \dodoi{10.1086/421702}

\bibitem[{{Smith} {et~al.}(2012){Smith}, {Foster}, \& {Brickhouse}}]{AtomDB}
{Smith}, R.~K., {Foster}, A.~R., \& {Brickhouse}, N.~S. 2012, Astronomische
  Nachrichten, 333, 301, \dodoi{10.1002/asna.201211673}

\bibitem[{{Strickland} {et~al.}(2000){Strickland}, {Heckman}, {Weaver}, \&
  {Dahlem}}]{Strickland2000}
{Strickland}, D.~K., {Heckman}, T.~M., {Weaver}, K.~A., \& {Dahlem}, M. 2000,
  \aj, 120, 2965, \dodoi{10.1086/316846}

\bibitem[{{Strickland} {et~al.}(2002){Strickland}, {Heckman}, {Weaver},
  {Hoopes}, \& {Dahlem}}]{Strickland2002}
{Strickland}, D.~K., {Heckman}, T.~M., {Weaver}, K.~A., {Hoopes}, C.~G., \&
  {Dahlem}, M. 2002, \apj, 568, 689, \dodoi{10.1086/338889}

\bibitem[{{Tashiro} {et~al.}(2018){Tashiro}, {Maejima}, {Toda}, {Kelley},
  {Reichenthal}, {Lobell}, {Petre}, {Guainazzi}, {Costantini}, {Edison},
  {Fujimoto}, {Grim}, {Hayashida}, {den Herder}, {Ishisaki}, {Paltani},
  {Matsushita}, {Mori}, {Sneiderman}, {Takei}, {Terada}, {Tomida}, {Akamatsu},
  {Angelini}, {Arai}, {Awaki}, {Babyk}, {Bamba}, {Barfknecht}, {Barnstable},
  {Bialas}, {Blagojevic}, {Bonafede}, {Brambora}, {Brenneman}, {Brown},
  {Brown}, {Burns}, {Canavan}, {Carnahan}, {Chiao}, {Comber}, {Corrales}, {de
  Vries}, {Dercksen}, {Diaz-Trigo}, {Dillard}, {DiPirro}, {Done}, {Dotani},
  {Ebisawa}, {Eckart}, {Enoto}, {Ezoe}, {Ferrigno}, {Fukazawa}, {Fujita},
  {Furuzawa}, {Gallo}, {Graham}, {Gu}, {Hagino}, {Hamaguchi}, {Hatsukade},
  {Hawes}, {Hayashi}, {Hegarty}, {Hell}, {Hiraga}, {Hodges-Kluck}, {Holland},
  {Hornschemeier}, {Hoshino}, {Ichinohe}, {Iizuka}, {Ishibashi}, {Ishida},
  {Ishikawa}, {Ishimura}, {James}, {Kallman}, {Kara}, {Katsuda}, {Kenyon},
  {Kilbourne}, {Kimball}, {Kitaguti}, {Kitamoto}, {Kobayashi}, {Kohmura},
  {Koyama}, {Kubota}, {Leutenegger}, {Lockard}, {Loewenstein}, {Maeda},
  {Marbley}, {Markevitch}, {Matsumoto}, {Matsuzaki}, {McCammon}, {McNamara},
  {Miko}, {Miller}, {Miller}, {Minesugi}, {Mitsuishi}, {Mizuno}, {Mori},
  {Mukai}, {Murakami}, {Mushotzky}, {Nakajima}, {Nakamura}, {Nakashima},
  {Nakazawa}, {Natsukari}, {Nigo}, {Nishioka}, {Nobukawa}, {Nobukawa}, {Noda},
  {Odaka}, {Ogawa}, {Ohashi}, {Ohno}, {Ohta}, {Okajima}, {Okamoto}, {Onizuka},
  {Ota}, {Ozaki}, {Plucinsky}, {Porter}, {Pottschmidt}, {Sato}, {Sato},
  {Sawada}, {Seta}, {Shelton}, {Shibano}, {Shida}, {Shidatsu}, {Shirron},
  {Simionescu}, {Smith}, {Someya}, {Soong}, {Suagawara}, {Szymkowiak},
  {Takahashi}, {Tamagawa}, {Tamura}, {Tanaka}, {Terashima}, {Tsuboi},
  {Tsujimoto}, {Tsunemi}, {Tsuru}, {Uchida}, {Uchiyama}, {Ueda}, {Uno},
  {Walsh}, {Watanabe}, {Williams}, {Wolfs}, {Wright}, {Yamada}, {Yamaguchi},
  {Yamaoka}, {Yamasaki}, {Yamauchi}, {Yamauchi}, {Yanagase}, {Yaqoob},
  {Yasuda}, {Yoshioka}, {Zabala}, \& {Irina}}]{XRISM}
{Tashiro}, M., {Maejima}, H., {Toda}, K., {et~al.} 2018, in Society of
  Photo-Optical Instrumentation Engineers (SPIE) Conference Series, Vol. 10699,
  Space Telescopes and Instrumentation 2018: Ultraviolet to Gamma Ray, ed.
  J.-W.~A. {den Herder}, S.~{Nikzad}, \& K.~{Nakazawa}, 1069922,
  \dodoi{10.1117/12.2309455}

\bibitem[{{Thompson} {et~al.}(2016){Thompson}, {Quataert}, {Zhang}, \&
  {Weinberg}}]{Thompson2016}
{Thompson}, T.~A., {Quataert}, E., {Zhang}, D., \& {Weinberg}, D.~H. 2016,
  \mnras, 455, 1830, \dodoi{10.1093/mnras/stv2428}

\bibitem[{{Veilleux} {et~al.}(2005){Veilleux}, {Cecil}, \&
  {Bland-Hawthorn}}]{Veilleux2005}
{Veilleux}, S., {Cecil}, G., \& {Bland-Hawthorn}, J. 2005, \araa, 43, 769,
  \dodoi{10.1146/annurev.astro.43.072103.150610}

\bibitem[{{Veilleux} {et~al.}(2020){Veilleux}, {Maiolino}, {Bolatto}, \&
  {Aalto}}]{Veilleux2020}
{Veilleux}, S., {Maiolino}, R., {Bolatto}, A.~D., \& {Aalto}, S. 2020, \aapr,
  28, 2, \dodoi{10.1007/s00159-019-0121-9}

\bibitem[{{Verner} {et~al.}(1996){Verner}, {Ferland}, {Korista}, \&
  {Yakovlev}}]{Verner1996}
{Verner}, D.~A., {Ferland}, G.~J., {Korista}, K.~T., \& {Yakovlev}, D.~G. 1996,
  \apj, 465, 487, \dodoi{10.1086/177435}

\bibitem[{{Wang}(1995)}]{Wang1995}
{Wang}, B. 1995, \apj, 444, 590, \dodoi{10.1086/175633}

\bibitem[{{Wang} \& {Liu}(2012)}]{Wang2012}
{Wang}, Q.~D., \& {Liu}, J. 2012, Astronomische Nachrichten, 333, 373,
  \dodoi{10.1002/asna.201211657}

\bibitem[{{Westmoquette} {et~al.}(2011){Westmoquette}, {Smith}, \&
  {Gallagher}}]{Westmoquette2011}
{Westmoquette}, M.~S., {Smith}, L.~J., \& {Gallagher}, J.~S., I. 2011, \mnras,
  414, 3719, \dodoi{10.1111/j.1365-2966.2011.18675.x}

\bibitem[{{Wik} {et~al.}(2014){Wik}, {Lehmer}, {Hornschemeier}, {Yukita},
  {Ptak}, {Zezas}, {Antoniou}, {Argo}, {Bechtol}, {Boggs}, {Christensen},
  {Craig}, {Hailey}, {Harrison}, {Krivonos}, {Maccarone}, {Stern}, {Venters},
  \& {Zhang}}]{Wik2014}
{Wik}, D.~R., {Lehmer}, B.~D., {Hornschemeier}, A.~E., {et~al.} 2014, \apj,
  797, 79, \dodoi{10.1088/0004-637X/797/2/79}

\end{thebibliography}



\end{document}